\documentclass[fleqn,usenatbib]{mnras}

\usepackage{newtxtext,newtxmath}

\usepackage[T1]{fontenc}
\usepackage{ae,aecompl}

\usepackage{graphicx}	
\usepackage{amsmath}	
\usepackage{amssymb}	
\usepackage{nccmath}

\newcommand\uclchem{{\sc uclchem}~}
\newcommand\formamide{NH$_2$CHO~}
\newcommand\methyl{CH$_3$NCO~}
\graphicspath{{./}{./images/}}

\let\oldbibitem\bibitem
\def\bibitem{\vskip 0.02cm\oldbibitem}

\title[Peptide-like complex organic molecules in star-forming regions]{Chemical modelling of complex organic molecules with peptide-like bonds in star-forming regions}

\author[D. Qu\'enard et al.]
{
David Qu\'enard,$^{1}$\thanks{E-mail: d.quenard@qmul.ac.uk}
Izaskun Jim\'enez-Serra,$^{1}$
Serena Viti,$^{2}$
Jonathan Holdship,$^{2}$
\newauthor and Audrey Coutens$^{3}$
\\
$^{1}$School of Physics and Astronomy, Queen Mary University of London, Mile End Road, London E1 4NS, UK\\
$^{2}$Department of Physics and Astronomy, University College London, Gower Street, London, WC1E 6BT, UK\\
$^{3}$Laboratoire d'astrophysique de Bordeaux, Univ. Bordeaux, CNRS, B18N, allée Geoffroy Saint-Hilaire, 33615 Pessac, France\\
}

\date{Accepted XXX. Received YYY; in original form ZZZ}

\pubyear{2017}

\begin{document}
\label{firstpage}
\pagerange{\pageref{firstpage}--\pageref{lastpage}}
\maketitle

\begin{abstract}
Peptide bonds (N-C=O) play a key role in metabolic processes since they link amino acids into peptide chains or proteins. Recently, several molecules containing peptide-like bonds have been detected across multiple environments in the interstellar medium (ISM), growing the need to fully understand their chemistry and their role in forming larger pre-biotic molecules. We present a comprehensive study of the chemistry of three molecules containing peptide-like bonds: HNCO, NH$_2$CHO, and CH$_3$NCO. We also included other CHNO isomers (HCNO, HOCN), and C$_2$H$_3$NO isomers (CH$_3$OCN, CH$_3$CNO) to the study. We have used the \uclchem gas-grain chemical code and included in our chemical network all possible formation/destruction pathways of these peptide-like molecules recently investigated either by theoretical calculations or in laboratory experiments. Our predictions are compared to observations obtained toward the proto-star IRAS16293--2422 and the L1544 pre-stellar core. Our results show that some key reactions involving the CHNO and C$_2$H$_3$NO isomers need to be modified to match the observations. Consistently with recent laboratory findings, hydrogenation is unlikely to produce NH$_2$CHO on grain surfaces, while a combination of radical-radical surface reactions and gas-phase reactions is a better alternative. In addition, better results are obtained for NH$_2$CHO when a slightly higher activation energy of 25$\,$K is considered for the gas-phase reaction $\rm NH_2 + H_2CO \rightarrow NH_2CHO + H$. Finally, our modelling shows that the observed correlation between NH$_2$CHO and HNCO in star-forming regions may come from the fact that HNCO and NH$_2$CHO react to temperature in the same manner rather than from a direct chemical link between the two species.
\end{abstract}

\begin{keywords}
Astrochemistry -- Molecular data -- ISM: abundances -- methods: numerical
\end{keywords}

\section{Introduction}
Complex organic molecules are molecules with more than six atoms in their structure containing at least one carbon atom \citep{herbst2009}. Among them, O-bearing and N-bearing COMs have attracted much interest in recent years because of their role in prebiotic chemistry \citep{saladino2012}. Numerous O-bearing and N-bearing COMs have been detected in the ISM and toward comets, suggesting a potential link between the two \citep[e.g.][]{crovisier2004,goesmann2015,altwegg2017}. N-bearing COMs containing peptide-like bonds (N-C=O bond) are of particular interest because of their role in linking amino acids into protein chains. Some peptide-like molecules are isocyanic acid (HNCO), formamide (NH$_2$CHO), and methyl isocyanate (CH$_3$NCO).


Isocyanic acid and formamide were first detected in space in early observations toward Sgr B2 \citep{rubin1971,snyder1972}, and since then they have been found in a variety of astrophysical environments \citep[see][and references therein]{turner1999, quan2010,lopez-sepulcre2015}. The stable isomers of HNCO, cyanic acid (HOCN) and fulminic acid (HCNO), have also been reported, respectively, in Sgr B2(OH) \citep{brunken2009} and in several cold sources (e.g. B1 and L1544) and the lukewarm environment of L1527 \citep{marcelino2009}.

Several efforts have been made recently to understand the chemistry of these species in various environments, especially for formamide for which its formation pathways are under strong debate. Indeed, observational and theoretical studies \citep{charnley1997, mendoza2014, lopez-sepulcre2015, song2016} initially proposed a scenario in which formamide formed through successive hydrogenation of HNCO on grain surfaces. However, laboratory experiments showed that hydrogenation is ineffective \citep{noble2015, fedoseev2015} and radical-radical reactions such as the one between NH$_2$ and H$_2$CO (or HCO) are instead favoured \citep{jones2011, fedoseev2016}. The equivalent gas phase formation reaction also seems to be a possible route, as shown by recent theoretical calculations presented by \citet{barone2015} and \citet{skouteris2017}.

In contrast to HNCO and NH$_2$CHO, methyl isocyanate has been discovered only recently in high-mass star-forming regions \citep[toward SgrB2 and Orion;][]{halfen2015,cernicharo2016}. The first detection of this molecule around a Solar-type proto-star (IRAS16293--2422; hereafter IRAS16293) has just been reported by \citet{martin-domenech2017} and \citet{ligterink2017}. CH$_3$NCO was proposed to be tentatively detected in the comet 67P/Churyumov--Gerasimenko by the Cometary Sampling and Composition (COSAC) mass spectrometer on board the \textit{Rosetta} spacecraft's Philae lander \citep{goesmann2015}, but this detection does not seem to be confirmed \citep{altwegg2017}. Several formation routes for this molecule (both in the gas-phase and on the grain surface) have been suggested \citep[][]{charnley1997,halfen2015,cernicharo2016,belloche2017}, but a comprehensive modelling of the chemistry of the C$_2$H$_3$NO isomers is still missing.

In this paper, we investigate the dominant formation pathways of these molecules in various relevant astrophysical environments (the IRAS16293 B hot corino and its cold envelope, and the L1544 pre-stellar core), by comparing the observed abundance of HNCO, NH$_2$CHO and CH$_3$NCO to the predicted one obtained by chemical modelling. 
In Section 2, we present the gas-grain chemical code \uclchem and we report the chemical network of the CHNO isomers, NH$_2$CHO and the C$_2$H$_3$NO isomers used in our modelling, for which we have collected all recent information available from laboratory experiments and theoretical calculations. In Section 3, we describe the sources to be modelled and Section 4 reports our modelling results. In Section 5 and 6, we discuss the formation and origin of formamide in various environments and Section 7 finally summarises our conclusions.

\section{Chemical code and network}

We have used the gas-grain chemical code \uclchem \citep{holdship2017} to model the chemistry of the peptide-like species HNCO, NH$_2$CHO and CH$_3$NCO in star-forming regions. The gas-phase reaction network of \uclchem is based on the UMIST database \citep{mcelroy2013} although additional gas-phase reactions have been included (see below). In the context of our work, \uclchem has been updated with a new treatment for grain-surface reactions that now considers the processes of grain surface diffusion \citep{hasegawa1992}, chemical reactive desorption \citep{minissale2016}, and reaction-diffusion competition \citep{chang2007, garrod2011, ruaud2016}. In Appendix \ref{chem_proc}, we describe how these processes have been implemented in the code.

Following this new treatment for grain-surface reactions in {\sc uclchem}, we have added reactions to hydrogenate C, N, and O atoms into their non-saturated and saturated forms. These hydrogenation processes can also now lead to chemical reactive desorption, releasing newly formed molecules in the gas phase. The full network contains 364 species (243 in the gas phase and 121 on the surface of the grains) and 3446 reactions. Binding energies of molecules have been taken from \citet{wakelam2017}. In the following, we describe the chemical reactions (both gas phase and grain surface) that have been added for HNCO and its isomers, NH$_2$CHO, and CH$_3$NCO and its isomers to the chemical network of {\sc uclchem}.

\subsection{HNCO and isomers}\label{network_hnco}

Isocyanic acid (HNCO) is the most stable isomer of the CHNO species. The other isomers are, in order of energy difference with respect to HNCO, cyanic acid (HOCN), fulminic acid (HCNO), and isofulminic acid (HONC). The chemistry of these molecules has been recently studied by \citet{quan2010} following the observations of HNCO, HCNO, HOCN toward several molecular dark clouds \citep[e.g.][]{turner1999, marcelino2009, marcelino2010}. A similar study of the chemistry of HNCO has been performed by \citet{tideswell2010}.
The gas-phase chemical network of \citet{quan2010} (see their Table 1) is already included in the UMIST database. We note, however, that we had to modify the parameters (reaction rate and/or activation energy barrier) for some of these gas-phase reactions in order to match the observations. We have also implemented in \uclchem their grain surface network, as presented in the Table 2 of \citet{quan2010}.

For HNCO we also added the following grain-surface reaction\footnote{In the following, ``$\#$'' means that the species is located on the grain surface.}:
\begin{ceqn}
	\begin{equation}
		\mathrm{\#NH + \#CO} \longrightarrow \mathrm{\#HNCO}\label{nh+co}
	\end{equation}
\end{ceqn}
\citet{belloche2017} investigated the energy barrier for this reaction (originally estimated to be 2500\,K by \citet{garrod2008}) by varying it in a grid models. They found that an efficient reaction (\ref{nh+co}) with a barrier of 1250\,K provided the best fit to their observations; however, they did not use the \citet{quan2010} network. This reaction has been extensively investigated by \citet{himmel2002} both experimentally and through quantum chemical calculations and they derive a barrier of 4200\,K. This value has been also used by \citet{fedoseev2015} and this is the one we consider in our work.

The complete chemical network is provided in Tables \ref{table_reac_grain} and \ref{table_reac_gas} and the modifications will be discussed in Section \ref{result_hnco}.

\subsection{\formamide}\label{network_formamide}

In the literature, two major grain surface formation routes of formamide (NH$_2$CHO) are proposed.
The first one is the successive hydrogenation of isocyanic acid, HNCO, on the grain surface:
\begin{ceqn}
	\begin{eqnarray}
		\mathrm{\#HNCO + \#H} &\longrightarrow& \mathrm{\#H_2NCO}\label{firstH}\\
		\mathrm{\#H_2NCO + \#H} &\longrightarrow& \mathrm{\#NH_2CHO}\label{secondH}
	\end{eqnarray}
\end{ceqn}
Reaction (\ref{secondH}) is barrierless since it is a radical-radical reaction but this is not the case for reaction (\ref{firstH}) ($E{\rm _A}=1962$\,K, see Table \ref{table_reac_grain}).
These reactions have been first proposed by \citet{charnley1997} and supported by \citet{raunier2004} following an experiment of UV irradiation of pure ices of HNCO at 10\,K. \citet{mendoza2014} and \citet{lopez-sepulcre2015} showed an observational correlation between the abundance of HNCO and \formamide for different star-forming regions spanning from low-mass to high-mass star forming regions, shock regions and cold envelopes (see their Fig. 2). In their works, they proposed that HNCO and \formamide are chemically linked: either both species are formed from the same precursor or one forms from the other. In their view, \formamide forms on the grain surface from successive hydrogenation of HNCO and is released when the temperature rises enough to sublimate the ices. This is consistent with the spatial correlation as well as the similar deuteration found in IRAS16293 for the two species \citep{coutens2016}. This theory has been supported by calculations performed by \citet{song2016} on the formation of \formamide on amorphous solid water (ASW) surfaces and in the gas phase. In particular, they show that quantum tunnelling of hydrogen atoms greatly increases the rate of reaction (\ref{firstH}). However, note that these calculations were performed down to 103\,K in the ASW surfaces, which is much larger than the cold conditions found in molecular dark clouds ($\sim$10$-$20\,K).

Recent laboratory experiments have however challenged this scenario \citep[see][]{noble2015,fedoseev2015}. \citet{noble2015} bombarded a pure HNCO ice with H atoms and they followed both the gas phase and grain surface composition. In their experiments, no formamide was produced within the detectable limits, which led them to conclude that HNCO does not saturate, and hence reaction (\ref{secondH}) does not lead to \formamide but to HNCO via:
\begin{ceqn}
	\begin{equation}
		\mathrm{\#H_2NCO + \#H} \longrightarrow \mathrm{\#HNCO + \#H_2}\label{noble_reac}
	\end{equation}
\end{ceqn}
This result was confirmed by similar experiments performed by \citet{fedoseev2015}.

The second grain-surface formation route involves the radical-radical reaction between NH$_2$ and HCO or H$_2$CO:
\begin{ceqn}
	\begin{eqnarray}
		\mathrm{\#NH_2 + \#HCO} &\longrightarrow& \mathrm{\#NH_2CHO}\label{with_HCO}\\
							&\longrightarrow& \mathrm{\#NH_3 + CO}\\
		\mathrm{\#NH_2 + \#H_2CO} &\longrightarrow& \mathrm{\#NH_2CHO + \#H}\label{with_H2CO}\\
							&\longrightarrow& \mathrm{\#NH_3 + \#HCO}
\end{eqnarray}
\end{ceqn}
This grain surface chemistry was proposed by \citet{fedoseev2016} after carrying out low temperature ($\sim$\,13\,K) laboratory experiments. In their work, they show that the UV irradiation of hydroxylamine (NH$_2$OH, formed from NO contained in CO-, H$_2$CO- and CH$_3$OH-rich ices at 13\,K) yields a significant amount of NH$_2$ and, subsequently, HNCO, OCN$^-$ and \formamide (see their Figure 7). This result may explain the observed correlation between HNCO and \formamide \citep{mendoza2014,lopez-sepulcre2015}, since both species may originate from the same precursor on the grain surface.

In the gas phase, a similar network was proposed by \citet{kahane2013} ensuing a study of the low-mass proto-star IRAS16293:
\begin{ceqn}
	\begin{equation}
		\mathrm{NH_2 + H_2CO} \longrightarrow \mathrm{NH_2CHO + H}\label{with_H2CO_gas}
	\end{equation}
\end{ceqn}
The gas phase pathway has been studied theoretically by \citet{barone2015} and \citet{skouteris2017}. \citet{barone2015} found that reaction (\ref{with_H2CO_gas}) has a small activation energy of 26.9\,K. However, in their computation, they did not consider the formation of a Van der Waals complex in the reaction path. This has been considered in \citet{skouteris2017} and they re-evaluated both the reaction rate and the activation energy of the reaction, now equal to 4.88\,K, i.e. much lower than the one found by \citet{barone2015}. One should note that \citet{song2016}, from their calculations, disputed the feasibility of reaction (\ref{with_H2CO_gas}) arguing that a significant barrier prevents the reaction from occurring. In Section \ref{result_formamide}, we will discuss the effects of these two different rates on the chemistry of NH$_2$CHO.

Other gas phase formation reactions involving NH$_2$CHO have been introduced in our network as, for instance, the one proposed by \citet{quan2007}:
\begin{ceqn}
	\begin{equation}
		\mathrm{NH_4^+ + H_2CO} \longrightarrow \mathrm{NH_4CH_2O^+ + h\nu}
	\end{equation}
\end{ceqn}
followed by dissociative recombination to form formamide. According to the same authors, this pathway is not efficient and does not produce much formamide.
\citet{halfen2011} also proposed a formation route, based on the previous one:
\begin{ceqn}
	\begin{eqnarray}
		\mathrm{NH_4^+ + H_2CO} &\longrightarrow& \mathrm{NH_3CHO^+ + H_2}\label{high_barrier}\\
		\mathrm{NH_3CHO^+ + e^-} &\longrightarrow& \mathrm{NH_2CHO + H}
	\end{eqnarray}
\end{ceqn}
Calculations performed by \citet{redondo2014-1} showed that reaction (\ref{high_barrier}) has a high barrier and is unlikely to occur. In a previous work, \citet{redondo2014} (see their Table 1) also studied the formation of formamide via the interaction of (NH$_3^+$, NH$_4^+$, NH$_3$OH$^+$, and NH$_2$OH$^+$) with (HCO, H$_2$CO, and HCOOH). Some of these reactions are:
\begin{ceqn}
	\begin{eqnarray}
		\mathrm{NH_3^+ + HCOOH} &\longrightarrow& \mathrm{NH_2CHO^+ + H_2O}\\
		\mathrm{NH_3OH^+ + HCO} &\longrightarrow& \mathrm{NH_2CHO^+ + H_2O}\\
		\mathrm{NH_2OH^+ + HCO} &\longrightarrow& \mathrm{NH_2CHO^+ + OH}\\
		\mathrm{NH_2OH^+ + H_2CO} &\longrightarrow& \mathrm{NH_2CHO^+ + H_2O}\\
		\mathrm{NH_2OH + H_2CO} &\longrightarrow& \mathrm{NH_2CHO + H_2O}
	\end{eqnarray}
\end{ceqn}
Their computational results show that all their proposed reactions are subject to a high barrier and are not feasible under interstellar conditions.

The destruction routes of formamide are poorly constrained. From the KIDA database\footnote{\url{http://kida.obs.u-bordeaux1.fr}} \citep{wakelam2012}, we have considered a set of destruction reactions of formamide with ions in the gas phase.
On the grain surface, formamide is destroyed following two reaction pathways:
\begin{ceqn}
	\begin{eqnarray}
		\mathrm{ \#NH_2CHO + \#OH} &\longrightarrow& \mathrm{\#H_2NCO + \#H_2O}\label{des_form_belloche}\\
		\mathrm{\#NH_2CHO + \#CH_2} &\longrightarrow& \mathrm{\#CH_3CONH_2}
	\end{eqnarray}
\end{ceqn}
Reaction (\ref{des_form_belloche}) has been discussed by \citet{belloche2017} where they show that H$_2$NCO is more likely to be produced than HNCHO, according to their best results. Thus, we did not include HNCHO in our study.

The complete set of grain surface and gas-phase reactions included in our study is shown in Tables \ref{table_reac_grain} and \ref{table_reac_gas}, respectively.

\subsection{\methyl and isomers}\label{network_methyl}

\citet{halfen2015} proposed a basic reaction network to form methyl isocyanate in space. It was mostly a gas phase chemistry where \methyl is mainly formed from methylation of HNCO:
\begin{ceqn}
	\begin{equation}
		\mathrm{CH_3 + HNCO} \longrightarrow \mathrm{CH_3NCO + H}\label{methyl_grain1}
	\end{equation}
\end{ceqn}

This pathway is also favoured by \citet{goesmann2015}. The grain surface route proposed by \citet{belloche2017} has also been introduced in our chemical network:
\begin{ceqn}
	\begin{equation}
		\mathrm{\#CH_3 + \#OCN} \longrightarrow \mathrm{\#CH_3NCO}
	\end{equation}
\end{ceqn}
This radical-radical reaction is barrierless and exothermic and thus should be efficient on the grain surface. This is confirmed by laboratory studies led by \citet{ligterink2017}. Their experiment also suggests that the grain-surface equivalent reaction to reaction (\ref{methyl_grain1}) occurs on  grain surfaces. 

Besides these formation reactions, we have added to the network a destruction route for \methyl on the grain surface (N. Ligterink, private communication):
\begin{ceqn}
	\begin{equation}
		\mathrm{\#CH_3NCO + \#H} \longrightarrow \mathrm{\#CH_3NH + \#CO}\label{ch3nco_h}
	\end{equation}
\end{ceqn}
Similar reactions have been also included for the different isomers.\\

\methyl is not the only stable isomer of C$_2$H$_3$NO: methyl cyanate (CH$_3$OCN), acetonitrile N-oxide (CH$_3$CNO), fulminate (CH$_3$ONC), and CH$_3$CON are all stable isomers, although the stability of CH$_3$CON is still under debate \citep{dalbouha2016}. The energy difference of these isomers, with respect to methyl isocyanate, is +109.8, +239.9, +347.8, and +618.0\,kJ\,mol$^{-1}$, respectively.
A preliminary chemistry study of these species has been performed by \citet{martin-domenech2017} for the physical conditions of the low-mass proto-star IRAS16293 B. A more complete network is reported here where the whole chemical network is built up based on the reactions of \methyl\!. We discuss the parameters of these reactions in Section \ref{result_methyl}.   
All the reactions added to \uclchem involving the C$_2$H$_3$NO isomers are reported in Tables \ref{table_reac_grain} and \ref{table_reac_gas} for the grain surface and the gas phase, respectively.

\begin{table}
	\centering
	\caption{grain-surface reactions related to \formamide and C$_2$H$_3$NO isomers.\label{table_reac_grain}}
	\begin{tabular}{lccc}
		\hline\hline
		Reaction	&	\multicolumn{2}{c}{Parameters}	& Ref.\\
		\hline
		Radical diffusion reactions				&	$E{\rm _b^A}$ (K)	&	$E{\rm _b^B}$ (K)	&\\
		\hline
		$\rm \#NH + \#CO \rightarrow \#HNCO$					&	1300	&	650	&	9,10\\
		$\rm \#H + \#HOCN \rightarrow \#H_2O + \#CN$			&	325	&	2200	&	1\\
		$\rm \#H + \#HOCN \rightarrow \#H_2OCN$				&	325	&	2200	&	1\\
		$\rm \#H + \#HCNO \rightarrow \#CH_2 + \#NO$			&	325	&	2200	&	1\\
		$\rm \#H + \#HCNO \rightarrow \#H_2CNO$				&	325	&	2200	&	1\\
		$\rm \#H + \#HNCO \rightarrow \#NH_2 + \#CO$			&	325	&	2200	&	2,3,5\\
		$\rm \#H + \#HNCO \rightarrow \#H_2NCO$				&	325	&	2200	&	2,3,4\\
		$\rm \#H + \#H_2NCO \rightarrow \#NH_2CHO$			&	325	&	2553	&	2\\
		$\rm \#H + \#H_2NCO \rightarrow \#HNCO + \#H_2$			&	325	&	2553	&	6\\
		$\rm \#NH_2 + \#HCO \rightarrow \#NH_3 + \#CO$			&	1600	&	1200	&	7\\
		$\rm \#NH_2 + \#HCO \rightarrow \#NH_2CHO$			&	1600	&	1200	&	7\\
		$\rm \#NH_2 + \#H_2CO \rightarrow \#NH_3 + \#HCO$		&	1600	&	2250	&	7\\
		$\rm \#NH_2 + \#H_2CO \rightarrow \#NH_2CHO + \#H$		&	1600	&	2250	&	7\\
		$\rm \#NH_2CHO + \#OH \rightarrow \#H_2NCO + \#H_2O$	&	3150	&	2300	&	8\\
		$\rm \#NH_2CHO + \#CH_2 \rightarrow \#CH_3CONH_2$	&	3150	&	700	&	1\\
		$\rm \#H_2NCO + \#CH_3 \rightarrow \#CH_3CONH_2$		&	2553	&	800	&	8\\
		$\rm \#CH_3 + \#OCN \rightarrow \#CH_3NCO$			&	800	&	1200	&	8,11\\
		$\rm \#CH_3 + \#OCN \rightarrow \#CH_3OCN$			&	800	&	1200	&	1\\
		$\rm \#CH_3 + \#CNO \rightarrow \#CH_3CNO$			&	800	&	1200	&	1\\
		$\rm \#CH_3 + \#CNO \rightarrow \#CH_3ONC$			&	800	&	1200	&	1\\
		$\rm \#CH_3 + \#HNCO \rightarrow \#CH_3NCO + \#H$		&	800	&	2200	&	11\\
		$\rm \#CH_3 + \#HNCO \rightarrow \#CH_3OCN + \#H$		&	800	&	2200	&	1\\
		$\rm \#CH_3 + \#HNCO \rightarrow \#CH_4 + \#OCN$		&	800	&	2200	&	11\\
		$\rm \#CH_3 + \#HOCN \rightarrow \#CH_3NCO + \#H$		&	800	&	2200	&	1\\
		$\rm \#CH_3 + \#HOCN \rightarrow \#CH_3OCN + \#H$		&	800	&	2200	&	1\\
		$\rm \#CH_3 + \#HOCN \rightarrow \#CH_4 + \#OCN$		&	800	&	2200	&	1\\		
		$\rm \#CH_3 + \#HCNO \rightarrow \#CH_3CNO + \#H$		&	800	&	2200	&	1\\
		$\rm \#CH_3 + \#HCNO \rightarrow \#CH_3ONC + \#H$		&	800	&	2200	&	1\\
		$\rm \#CH_3 + \#HCNO \rightarrow \#CH_4 + \#CNO$		&	800	&	2200	&	1\\		
		$\rm \#CH_3 + \#HONC \rightarrow \#CH_3CNO + \#H$		&	800	&	2200	&	1\\
		$\rm \#CH_3 + \#HONC \rightarrow \#CH_3ONC + \#H$		&	800	&	2200	&	1\\
		$\rm \#CH_3 + \#HONC \rightarrow \#CH_4 + \#CNO$		&	800	&	2200	&	1\\
		$\rm \#H + \#CH_3NCO \rightarrow \#CH_3NH + \#CO$		&	325	&	2350	&	1\\
		$\rm \#H + \#CH_3NCO \rightarrow \#CH_3 + HNCO$		&	325	&	2350	&	1\\
		$\rm \#H + \#CH_3OCN \rightarrow \#CH_3OH + \#CN$		&	325	&	2350	&	1\\
		$\rm \#H + \#CH_3OCN \rightarrow \#CH_3 + \#HOCN$		&	325	&	2350	&	1\\
		$\rm \#H + \#CH_3CNO \rightarrow \#CH_3CN + \#OH$		&	325	&	2350	&	1\\
		$\rm \#H + \#CH_3CNO \rightarrow \#CH_3 + \#HCNO$		&	325	&	2350	&	1\\
		$\rm \#H + \#CH_3ONC \rightarrow \#CH_3OH + \#CN$		&	325	&	2350	&	1\\
		$\rm \#H + \#CH_3ONC \rightarrow \#CH_3 + \#HONC$		&	325	&	2350	&	1\\
		\hline
		Chemical reactive desorption		&	$E{\rm _D}$ (K)		&	$E{\rm _A}$ (K)	&\\
		\hline
		$\rm \#NH + \#CO \rightarrow HNCO$					&	4400	&	4200	&	9,10\\
		$\rm \#H + \#HOCN \rightarrow H_2O + CN$				&	5600	&	2300	&	1\\
		$\rm \#H + \#HOCN \rightarrow H_2OCN^+$				&	5106	&	1962	&	1\\
		$\rm \#H + \#HCNO \rightarrow CH_2 + NO$				&	1600	&	2300	&	1\\
		$\rm \#H + \#HCNO \rightarrow H_2CNO^+$				&	5106	&	1962	&	1\\
		$\rm \#H + \#HNCO \rightarrow NH_2 + CO$				&	3200	&	2300	&	2,3,5\\
		$\rm \#H + \#HNCO \rightarrow H_2NCO^+$				&	5106	&	1962	&	2,3,4\\
		$\rm \#H + \#H_2NCO \rightarrow NH_2CHO$				&	6300	&	0	&	2\\
		$\rm \#H + \#H_2NCO \rightarrow HNCO + H_2$			&	4400	&	0	&	6\\
		$\rm \#NH_2 + \#HCO \rightarrow NH_3 + CO$				&	5500	&	0	&	7\\
		$\rm \#NH_2 + \#HCO \rightarrow NH_2CHO$				&	6300	&	0	&	7\\
		$\rm \#NH_2 + \#H_2CO \rightarrow NH_3 + HCO$			&	5500	&	0	&	7\\
		$\rm \#NH_2 + \#H_2CO \rightarrow NH_2CHO$			&	6300	&	0	&	7\\
		$\rm \#NH_2CHO + \#OH \rightarrow H_2NCO^+ + H_2O$	&	5106	&	591	&	8\\
		$\rm \#NH_2CHO + \#CH_2 \rightarrow CH_3CONH_2$		&	6281	&	0	&	8\\
		$\rm \#H_2NCO + \#CH_3 \rightarrow CH_3CONH_2$		&	6281	&	0	&	8\\
		$\rm \#CH_3 + \#OCN \rightarrow CH_3NCO$				&	4700	&	0	&	8,11\\
		$\rm \#CH_3 + \#OCN \rightarrow CH_3OCN$				&	4700	&	0	&	1\\
		\end{tabular}
\end{table}

\begin{table}
	\centering
	\contcaption{grain-surface reactions related to \formamide and C$_2$H$_3$NO isomers.}
	\begin{tabular}{lccc}
		\hline\hline
		Reaction	&	\multicolumn{2}{c}{Parameters}	& Ref.\\
		\hline
		Chemical reactive desorption		&	$E{\rm _D}$ (K)		&	$E{\rm _A}$ (K)	&\\
		\hline
		$\rm \#CH_3 + \#CNO \rightarrow CH_3CNO$				&	4700	&	0	&	1\\
		$\rm \#CH_3 + \#CNO \rightarrow CH_3ONC$				&	4700	&	0	&	1\\
		$\rm \#CH_3 + \#HNCO \rightarrow CH_3NCO + H$			&	4700	&	0	&	11\\
		$\rm \#CH_3 + \#HNCO \rightarrow CH_3OCN + H$			&	4700	&	0	&	1\\
		$\rm \#CH_3 + \#HNCO \rightarrow CH_4 + OCN$			&	2400	&	0	&	11\\
		$\rm \#CH_3 + \#HOCN \rightarrow CH_3NCO + H$			&	2400	&	0	&	1\\
		$\rm \#CH_3 + \#HOCN \rightarrow CH_3OCN + H$			&	2400	&	0	&	1\\
		$\rm \#CH_3 + \#HOCN \rightarrow CH_4 + OCN$			&	2400	&	0	&	1\\		
		$\rm \#CH_3 + \#HCNO \rightarrow CH_3CNO + H$			&	2400	&	0	&	1\\
		$\rm \#CH_3 + \#HCNO \rightarrow CH_3ONC + H$			&	2400	&	0	&	1\\
		$\rm \#CH_3 + \#HCNO \rightarrow CH_4 + CNO$			&	2400	&	0	&	1\\		
		$\rm \#CH_3 + \#HONC \rightarrow CH_3CNO + H$			&	2400	&	0	&	1\\
		$\rm \#CH_3 + \#HONC \rightarrow CH_3ONC + H$			&	2400	&	0	&	1\\
		$\rm \#CH_3 + \#HONC \rightarrow CH_4 + CNO$			&	2400	&	0	&	1\\
		$\rm \#H + \#CH_3NCO \rightarrow CH_3NH + CO$			&	4681	&	0	&	1\\
		$\rm \#H + \#CH_3NCO \rightarrow CH_3 + HNCO$			&	4400	&	0	&	1\\
		$\rm \#H + \#CH_3OCN \rightarrow CH_3OH + CN$			&	5000	&	0	&	1\\
		$\rm \#H + \#CH_3OCN \rightarrow CH_3 + HOCN$			&	4400	&	0	&	1\\
		$\rm \#H + \#CH_3CNO \rightarrow CH_3CN + OH$			&	4680	&	0	&	1\\
		$\rm \#H + \#CH_3CNO \rightarrow CH_3 + HCNO$			&	4400	&	0	&	1\\
		$\rm \#H + \#CH_3ONC \rightarrow CH_3OH + CN$			&	5000	&	0	&	1\\
		$\rm \#H + \#CH_3ONC \rightarrow CH_3 + HONC$			&	4400	&	0	&	1\\
		\hline
		Direct cosmic-ray desorption		&									&\\
		H$_2$ formation induced desorption	&	\multicolumn{2}{c}{$E{\rm _D}$ (K)}		&\\
		Cosmic-ray induced UV desorption	&									&\\
		\hline
		$\rm \#HNCO \rightarrow HNCO$				&	\multicolumn{2}{c}{4400}	&	1\\
		$\rm \#HOCN \rightarrow HOCN$				&	\multicolumn{2}{c}{4400}	&	1\\
		$\rm \#HCNO \rightarrow HCNO$				&	\multicolumn{2}{c}{4400}	&	1\\
		$\rm \#HONC \rightarrow HONC$				&	\multicolumn{2}{c}{4400}	&	1\\
		$\rm \#H_2NCO \rightarrow H_2NCO^+$			&	\multicolumn{2}{c}{5106}	&	1\\
		$\rm \#H_2CNO \rightarrow H_2CNO^+$			&	\multicolumn{2}{c}{5106}	&	1\\
		$\rm \#H_2OCN \rightarrow H_2OCN^+$			&	\multicolumn{2}{c}{5106}	&	1\\
		$\rm \#HNCOH \rightarrow HNCOH^+$			&	\multicolumn{2}{c}{5106}	&	1\\
		$\rm \#HCNOH \rightarrow HCNOH^+$			&	\multicolumn{2}{c}{5106}	&	1\\
		$\rm \#NH_2CHO \rightarrow NH_2CHO$			&	\multicolumn{2}{c}{6300}	&	1\\
		$\rm \#CH_3NCO \rightarrow CH_3NCO$			&	\multicolumn{2}{c}{4700}	&	1\\
		$\rm \#CH_3OCN \rightarrow CH_3OCN$			&	\multicolumn{2}{c}{4700}	&	1\\
		$\rm \#CH_3CNO \rightarrow CH_3CNO$			&	\multicolumn{2}{c}{4700}	&	1\\
		$\rm \#CH_3ONC \rightarrow CH_3ONC$			&	\multicolumn{2}{c}{4700}	&	1\\
		$\rm \#CH_3CONH_2 \rightarrow CH_3CONH_2$	&	\multicolumn{2}{c}{6281}	&	1\\
		\hline
		\multicolumn{4}{l}{\textbf{References.} (1) estimation based on analogous reactions;}\\
		\multicolumn{4}{l}{(2) \citet{song2016}; (3) \citet{garrod2013};}\\
		\multicolumn{4}{l}{(4) \citet{nguyen1996}; (5) \citet{tsang1992};}\\
		\multicolumn{4}{l}{(6) \citet{noble2015}; (7) \citet{fedoseev2016};}\\
		\multicolumn{4}{l}{(8) \citet{belloche2017}; (9) \citet{fedoseev2015};}\\
		\multicolumn{4}{l}{(10) \citet{himmel2002}; (11) \citet{ligterink2017}.}
		\end{tabular}
\end{table}

\begin{table}
	\scriptsize
	\centering
	\caption{gas-phase reactions related to CHNO isomers, \formamide and C$_2$H$_3$NO isomers.\label{table_reac_gas}}
	\begin{tabular}{lcccc}
		\hline\hline
		Reaction	&	$\alpha$	&	$\beta$	&	$\gamma$	& Ref.\\
		\hline
		\multicolumn{5}{c}{CHNO\!}\\
		\hline
		$\rm CH_2 + NO \rightarrow HCNO$$^*$					&	$2.00(-12)$	&	$0$			&	$0$			&	1,6\\
		$\rm HOCN + O \rightarrow OH + OCN$$^*$				&	$3.33(-10)$	&	$0$			&	$195$		&	1,6\\
		$\rm HCNO + O \rightarrow CO + HNO$$^*$				&	$3.33(-10)$	&	$0$			&	$195$		&	1,6\\
		$\rm HONC + O \rightarrow OH + CNO$$^*$				&	$3.33(-10)$	&	$0$			&	$195$		&	1,6\\
		$\rm H_2OCN^+ + e^- \rightarrow HOCN + H$$^*$			&	$5.00(-09)$	&	$-0.50$		&	$0$			&	1,6\\
		$\rm HNCOH^+ + e^- \rightarrow HOCN + H$$^*$			&	$5.00(-09)$	&	$-0.50$		&	$0$			&	1,6\\
		\hline
		\multicolumn{5}{c}{\formamide\!}\\
		\hline
		$\rm NH_2 + H_2CO \rightarrow NH_2CHO$			&	$7.79(-15)$	&	$-2.56$		&	$25.0$		&	1,3,4\\
		$\rm NH_2CHO + CRPHOT \rightarrow NH_2 + H_2CO$	&	$6.00(+03)$	&	$0$			&	$0$			&	2\\
		$\rm H^+ + NH_2CHO \rightarrow HCO^+ + NH_3$		&	$2.50(-01)$	&	$4.82(-09)$	&	$6.62$		&	2\\
		$\rm H^+ + NH_2CHO \rightarrow HCNH^+ + H_2O$	&	$2.50(-01)$	&	$4.82(-09)$	&	$6.62$		&	2\\
		$\rm H^+ + NH_2CHO \rightarrow NH_2^+ + H_2CO$	&	$2.50(-01)$	&	$4.82(-09)$	&	$6.62$		&	2\\
		$\rm H^+ + NH_2CHO \rightarrow NH_4^+ + CO$		&	$2.50(-01)$	&	$4.82(-09)$	&	$6.62$		&	2\\
		$\rm He^+ + NH_2CHO \rightarrow He + NH_3 + CO^+$	&	$2.00(-01)$	&	$2.49(-09)$	&	$6.62$		&	2\\
		$\rm He^+ + NH_2CHO \rightarrow He + NH_3^+ + CO$	&	$2.00(-01)$	&	$2.49(-09)$	&	$6.62$		&	2\\
		$\rm He^+ + NH_2CHO \rightarrow He + NH + H_2CO^+$&	$2.00(-01)$	&	$2.49(-09)$	&	$6.62$		&	2\\
		$\rm He^+ + NH_2CHO \rightarrow He + NH^+ + H_2CO$&	$2.00(-01)$	&	$2.49(-09)$	&	$6.62$		&	2\\
		$\rm He^+ + NH_2CHO \rightarrow He + NH_4^+ CO$	&	$2.00(-01)$	&	$2.49(-09)$	&	$6.62$		&	2\\
		$\rm C^+ + NH_2CHO \rightarrow CH_3CN^+ + O$		&	$1.67(-01)$	&	$1.55(-09)$	&	$6.62$		&	2\\
		$\rm C^+ + NH_2CHO \rightarrow CH_2CN^+ + OH$	&	$1.67(-01)$	&	$1.55(-09)$	&	$6.62$		&	2\\
		$\rm C^+ + NH_2CHO \rightarrow CH_3CO^+ + N$		&	$1.67(-01)$	&	$1.55(-09)$	&	$6.62$		&	2\\
		$\rm C^+ + NH_2CHO \rightarrow HCN + H_2CO^+$	&	$1.67(-01)$	&	$1.55(-09)$	&	$6.62$		&	2\\
		$\rm C^+ + NH_2CHO \rightarrow HCN^+ + H_2CO$	&	$1.67(-01)$	&	$1.55(-09)$	&	$6.62$		&	2\\
		$\rm C^+ + NH_2CHO \rightarrow CN + H_2COH^+$	&	$1.67(-01)$	&	$1.55(-09)$	&	$6.62$		&	2\\
		$\rm C^+ + NH_2CHO \rightarrow CN + H_2COH^+$	&	$1.67(-01)$	&	$1.55(-09)$	&	$6.62$		&	2\\
		\hline
		\multicolumn{5}{c}{C$_2$H$_3$NO}\\
		\hline
		$\rm HNCO + CH_3 \rightarrow CH_3NCO + H$		&	$5.00(-11)$	&	$0$		&	$0$	&	1,5\\
		$\rm HOCN + CH_3 \rightarrow CH_3NCO + H$		&	$1.00(-20)$	&	$0$		&	$0$	&	1,5\\
		$\rm HNCO + CH_3 \rightarrow CH_3OCN + H$		&	$1.00(-20)$	&	$0$		&	$0$	&	1\\
		$\rm HOCN + CH_3 \rightarrow CH_3OCN + H$		&	$5.00(-11)$	&	$0$		&	$0$	&	1\\
		$\rm HCNO + CH_3 \rightarrow CH_3CNO + H$		&	$5.00(-11)$	&	$0$		&	$0$	&	1\\
		$\rm HONC + CH_3 \rightarrow CH_3CNO + H$		&	$1.00(-20)$	&	$0$		&	$0$	&	1\\
		$\rm HCNO + CH_3 \rightarrow CH_3ONC + H$		&	$1.00(-20)$	&	$0$		&	$0$	&	1\\
		$\rm HONC + CH_3 \rightarrow CH_3ONC + H$		&	$5.00(-11)$	&	$0$		&	$0$	&	1\\
		$\rm HNCO + CH_5^+ \rightarrow CH_3NCOH^+ + H_2$	&	$1.00(-09)$	&	$0$		&	$0$	&	5\\
		$\rm HOCN + CH_5^+ \rightarrow CH_3NCOH^+ + H_2$	&	$1.00(-09)$	&	$0$		&	$0$	&	5\\
		$\rm HCNO + CH_5^+ \rightarrow CH_3CNOH^+ + H_2$	&	$1.00(-09)$	&	$0$		&	$0$	&	1\\
		$\rm HONC + CH_5^+ \rightarrow CH_3CNOH^+ + H_2$	&	$1.00(-09)$	&	$0$		&	$0$	&	1\\
		$\rm CH_3NCOH^+ + e^- \rightarrow CH_3NCO + H$	&	$1.50(-07)$	&	$-0.50$	&	$0$	&	1,5\\
		$\rm CH_3NCOH^+ + e^- \rightarrow CH_3 + HOCN$	&	$1.50(-07)$	&	$-0.50$	&	$0$	&	1\\
		$\rm CH_3CNOH^+ + e^- \rightarrow CH_3CNO + H$	&	$1.50(-07)$	&	$-0.50$	&	$0$	&	1\\
		$\rm CH_3CNOH^+ + e^- \rightarrow CH_3 + HONC$	&	$1.50(-07)$	&	$-0.50$	&	$0$	&	1\\		
		\hline
		\multicolumn{5}{l}{\textbf{Notes.} $a(b)$ means $a\times10^b$.}\\
		\multicolumn{5}{l}{Bimolecular rate coefficients are tabulated as $k(T) = \alpha\left(\frac{T}{300}\right)^\beta\exp\left({-\frac{\gamma}{T}}\right)$}\\
		\multicolumn{5}{l}{in units of cm$^3$\,s$^{-1}$. Rate coefficients for cosmic-ray-induced photo-}\\
		\multicolumn{5}{l}{dissociation (CRPHOT) are tabulated in terms of $\zeta$ (s$^{-1}$). Reactions}\\
		\multicolumn{5}{l}{taken from the KIDA database are using the rate coefficient for ion-polar}\\
		\multicolumn{5}{l}{systems labelled as \textit{ionpol1} in KIDA: $k(T) = \alpha\beta\left[0.62 + 0.4767\,\gamma\left(\frac{300}{T}\right)^{0.5}\right]$.}\\
		\multicolumn{5}{l}{\textbf{References.} (1) this work; (2) KIDA database, \citet{wakelam2012};}\\
		\multicolumn{5}{l}{(3) \citet{barone2015}; (4) \citet{skouteris2017};}\\
		\multicolumn{5}{l}{(5) \citet{halfen2015}; (6) \citet{quan2010}.}
	\end{tabular}
\end{table}

\section{Chemical modelling and source selection}

\uclchem is run in three steps or phases:

\textit{Ambient cloud phase:} The first phase (referred to as Phase 0) computes the evolution of the chemistry in a diffuse cloud for 10$^6$ years. The initial gas phase atomic abundances (with respect to the total proton density n$_\textrm{H}$) are given in Table \ref{init_abund}. These values are taken from \citet[][case EA1]{wakelam2008} and are consistent with recent studies performed toward L1544 \citep{vasyunin2017, quenard2017-1} and IRAS16293 \citep{hincelin2011, bottinelli2014}. Such atomic abundances should be suitable initial conditions in cold dark clouds where heavy elements are depleted on dust grains. The initial density is kept constant to n$_\textrm{H}=10^2$\,cm$^{-3}$, as suggested by a similar chemical study of the pre-stellar core L1544 by \citet{quenard2017-1}. The temperature is set to $\rm T=10$\,K. The visual extinction is set to A$\rm _V=2$\,mag, corresponding to a size of $\sim$10\,pc. These are typical values for diffuse clouds.\\

\textit{Pre-stellar phase:} The second phase (referred to as Phase 1) uses the abundances obtained at the end of Phase 0 as initial conditions. In this step, a fragment of the cloud is slowly contracting, forming a pre-stellar core. The core contraction follows free-fall collapse \citep[as described in][]{rawlings1992} until it reaches the final density of the source (see Table \ref{phys_param}). The temperature does not change and stays the same throughout Phase 1 (i.e. $\rm T=10$\,K).\\

\textit{Proto-stellar phase:} The third phase (referred to as Phase 2) starts with the abundances derived at the end of Phase 1. In Phase 2, the temperature increases (keeping the density constant) following a rate defined by \citet{viti2004} for hot cores and by \citet{awad2010} for hot corinos. The final temperature depends on the source studied (see Table \ref{phys_param}). \\

\begin{table} 
	\centering
	\caption{Initial gas phase elemental abundances assumed relative to the total nuclear hydrogen density n$_{\textrm{H}}$.\label{init_abund}}
	\begin{tabular}{lcc}
		\hline\hline
		Species	&	EA1\\
		\hline
		He		&	$1.40\times10^{-1}$\\
		N		&	$2.14\times10^{-5}$\\
		O		&	$1.76\times10^{-4}$\\
		C$^+$	&	$7.30\times10^{-5}$\\
		S$^+$	&	$8.00\times10^{-8}$\\
		Si$^+$	&	$8.00\times10^{-9}$\\
		Mg$^+$	&	$7.00\times10^{-9}$\\
		Cl$^+$	&	$1.00\times10^{-9}$\\
		P$^+$	&	$2.00\times10^{-10}$\\
		F$^+$	&	$6.68\times10^{-9}$\\
		\hline
		\multicolumn{3}{l}{\textbf{Notes.} EA1 refers to the first elemental abundance model}\\
		\multicolumn{3}{l}{considered by \citet{wakelam2008}.}
	\end{tabular}
\end{table}

\begin{table} 
	\centering
	\caption{Temperature, density and visual extinction of the different sources modelled with \uclchem at the end of each phase.\label{phys_param}}
	\begin{tabular}{lccc}
		\hline\hline
		Source					&	Temperature	&	H density			&	A$\rm _V$\\
								&	(K)			&	(cm$^{-3}$)		&	(mag)\\
		\hline
		\multicolumn{3}{c}{Phase 0}\\
		\hline
		All sources				&	10			&	$1\times10^{2}$	&	2\\
		\hline
		\multicolumn{3}{c}{Phase 1}\\
		\hline
		L1544 (core centre)			&	10			&	$5\times10^{6}$	&	100\\
		L1544 (methanol peak)		&	10			&	$4\times10^{5}$	&	8\\
		IRAS16293 B (hot corino)		&	10			&	$5\times10^{8}$	&	1000\\
		IRAS16293 (cold envelope)	&	10			&	$2\times10^{6}$	&	30\\
		\hline
		\multicolumn{3}{c}{Phase 2}\\
		\hline
		IRAS16293 B (hot corino)		&	250			&	$5\times10^{8}$	&	1000\\
		IRAS16293 (cold envelope)	&	20			&	$2\times10^{6}$	&	30\\
		\hline
	\end{tabular}
\end{table}

Photo-processes and cosmic rays are set for all the different phases so that the chemistry depends on the external radiation field (G$_0=1$\,Habing) and the cosmic ray ionisation rate ($\zeta=1.3\times10^{-17}$\,s$^{-1}$). 

In order to test our chemical network, we have selected a few representative astrophysical sources to cover different types of physical and chemical environments: the pre-stellar core L1544 (core centre and methanol peak; see details below), the hot corino IRAS16293 B and its cold envelope. These sources are well-studied and many observations using both single-dish telescopes (e.g. IRAM 30m) and interferometers such as ALMA exist. In the following we describe the physical properties of the sources selected for our study:\\

\noindent \textbf{a) The pre-stellar core L1544.}\\

\noindent L1544 is a proto-typical pre-stellar core, which has been extensively studied in the past \citep[e.g.][]{caselli2002, caselli2003, quenard2016, quenard2017-1}.
The 1D physical structure of the source has been derived by \citet{keto2014} and it has been used to derive the values shown in Table \ref{phys_param}. In our modelling, we consider the position of the core centre, where most species are frozen out onto dust grains, and the position of the methanol peak reported by \citet{bizzocchi2014} and studied by \citet{jimenez-serra2016}. The core centre is characterised by n(H$_2$)$\sim5\times10^{6}$\,cm$^{-3}$ and large extinction (A$\rm _V\sim60$\,mag).

The methanol peak is detected $\sim$4000\,au away from the core centre. This ``methanol peak'' position reveals an enhancement of several COMs, as shown by \citet{vastel2014} and \citet{jimenez-serra2016}. This position has a density of a few 10$^{5}$\,cm$^{-3}$ and a moderate visual extinction \citep[A$\rm _V\sim$7$-$8\,mag within the core and not along the line-of-sight; see][]{jimenez-serra2016}. In our study, we will compare the molecular richness in these two different environments.\\

\noindent \textbf{b) The proto-star IRAS16293.}\\

\noindent IRAS16293 is a class 0 low-mass proto-star located in the $\rho$ Ophiuchus star-forming region, $147.3\pm3.4$\,pc away from us \citep{ortiz-leon2017}. IRAS16293 is at least a binary system \citep{wootten1989, pech2010}, with sources A and B separated by 5 arcsec (or $\sim$740\,au at 147.3\,pc).

Source B has been the object of numerous observations due to its chemical complexity \citep[e.g.][]{coutens2016, jorgensen2016}. Among the molecules detected in this source, we find methyl isocyanate which has been recently reported toward this object \citep{martin-domenech2017,ligterink2017}. Both studies give detection limits for many isomers of C$_2$H$_3$NO and CHNO.

We have used the 1D physical structure of IRAS16293 determined by \citet{crimier2010} to infer the physical conditions of its hot corino and cold envelope (see Table \ref{phys_param}). The parameters selected for the cold envelope are also consistent with previous modelling of this source by \citet{barone2015}. We have used the same chemical network as for L1544 to model the abundances of the same set of molecules for both the IRAS16293 B hot corino and the cold envelope surrounding it.

\section{Results}

In this section we will present the results of the chemical modelling for HNCO and its isomers (Section \ref{result_hnco}), CH$_3$NCO and its isomers (Section \ref{result_methyl}), and NH$_2$CHO (Section \ref{result_formamide}). Each section discusses the results for each of the four environments considered in our study (L1544 core centre and methanol peak and IRAS16293 hot corino and cold envelope) and the impact of the network related to each species on these results (e.g. key reactions, modifications). 

\subsection{HNCO and isomers}\label{result_hnco}

As explained in Section \ref{network_hnco}, the CHNO gas-grain network is the one proposed by \citet{quan2010}. The modelling using the original network of \citet{quan2010}, however, did not match the observations neither toward IRAS16293 nor toward L1544 (we stress that this was already noted by \cite{quan2010} for HOCN and HCNO toward L1544; see their Figure 6). Indeed, with the original network, HOCN and HCNO are overestimated by a large factor ($\sim$1000 and $\sim$30, respectively). Therefore, we needed to perform some changes to six key gas-phase reactions involving the formation and destruction of these two species (see reactions with an $^{*}$ in Table \ref{table_reac_gas}). Since the study of HOCN and HCNO is not the main focus of this work, we just investigate which reaction rates should be modified - and by how much - to obtain a better result. We however note that we did not modify any reaction involving HNCO, the most important isomer in our study.

In our chemical network, we increased the destruction rates of $\rm HOCN/HCNO\,+\,O$ from $3.33\times10^{-11}$ to $3.33\times10^{-10}$\,cm$^3$\,s$^{-1}$. We note that these reaction rates and energy barriers were just educated guesses assumed by \citet{quan2010}, and no experimental results do exist yet. In the case of HOCN, we also changed its barrier of 2470\,K to be the same as the reaction with HCNO (195\,K). For the two key gas-phase formation routes
\begin{ceqn}
	\begin{equation}
		\mathrm{H_2OCN^+/HNCOH^+} \longrightarrow \mathrm{HOCN + H},
	\end{equation}
\end{ceqn}
\noindent
we decreased the reaction rate from $1.50\times10^{-7}$ and $1.00\times10^{-7}$\,cm$^3$\,s$^{-1}$, respectively, to $5.00\times10^{-9}$\,cm$^3$\,s$^{-1}$. These dissociative recombination reactions were first introduced by \citet{marcelino2010} and their rates are poorly constrained \citep{quan2010}.

For HCNO, we found that reaction $\rm CH_2 + NO \rightarrow HCNO$ is the main formation route of HCNO. We decreased its reaction rate from $3.65\times10^{-11}$ to $2.00\times10^{-12}$\,cm$^3$\,s$^{-1}$. The latter reaction has been investigated theoretically and experimentally but the branching ratios for the out-coming products are still debated. For some studies HCNO is the dominant product \citep{glarborg1998,fikri2001,eshchenko2002} while for others it is not \citep{zhang2004}. \citet{marcelino2010} showed that this reaction can vary the final abundance of HCNO by several orders of magnitude. Although the conclusion obtained by \citet{zhang2004} is not supported by many other authors, it shows that this reaction is still poorly constrained.\\

\begin{figure*}
	\centering
	\includegraphics[width=0.49\hsize,clip=true,trim=0 0 0 0]{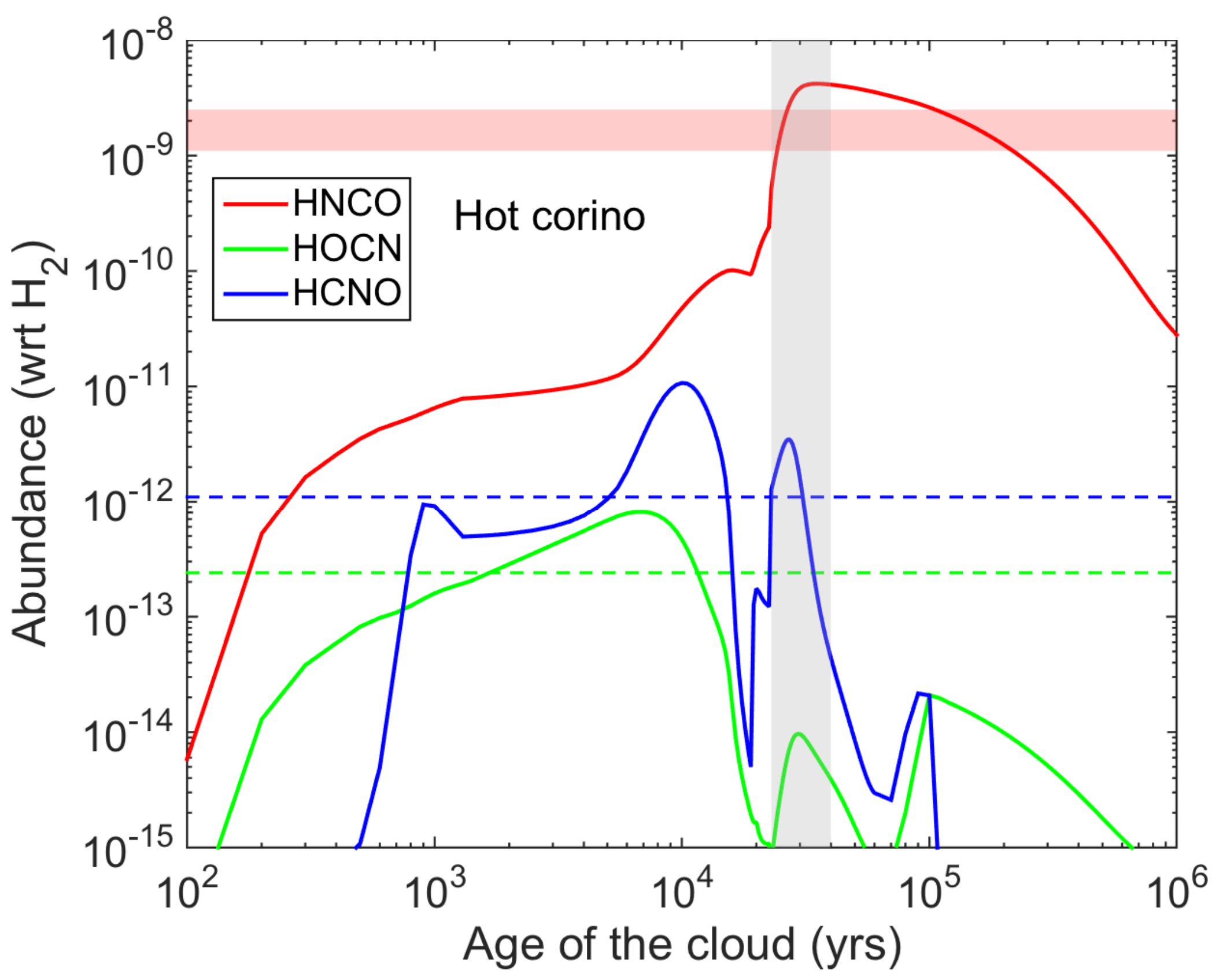}
	\includegraphics[width=0.49\hsize,clip=true,trim=0 0 0 0]{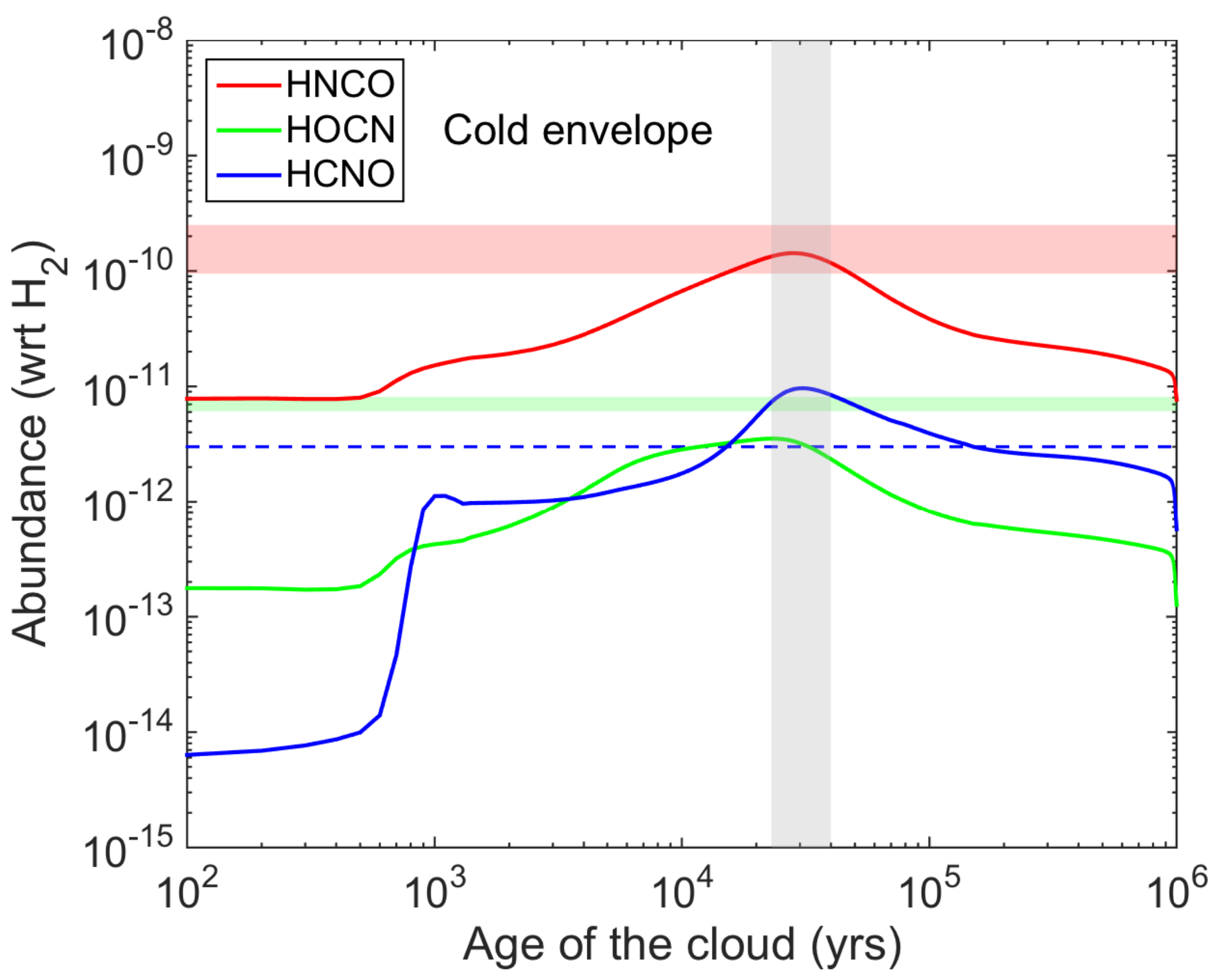}\\
	\includegraphics[width=0.49\hsize,clip=true,trim=0 0 0 0]{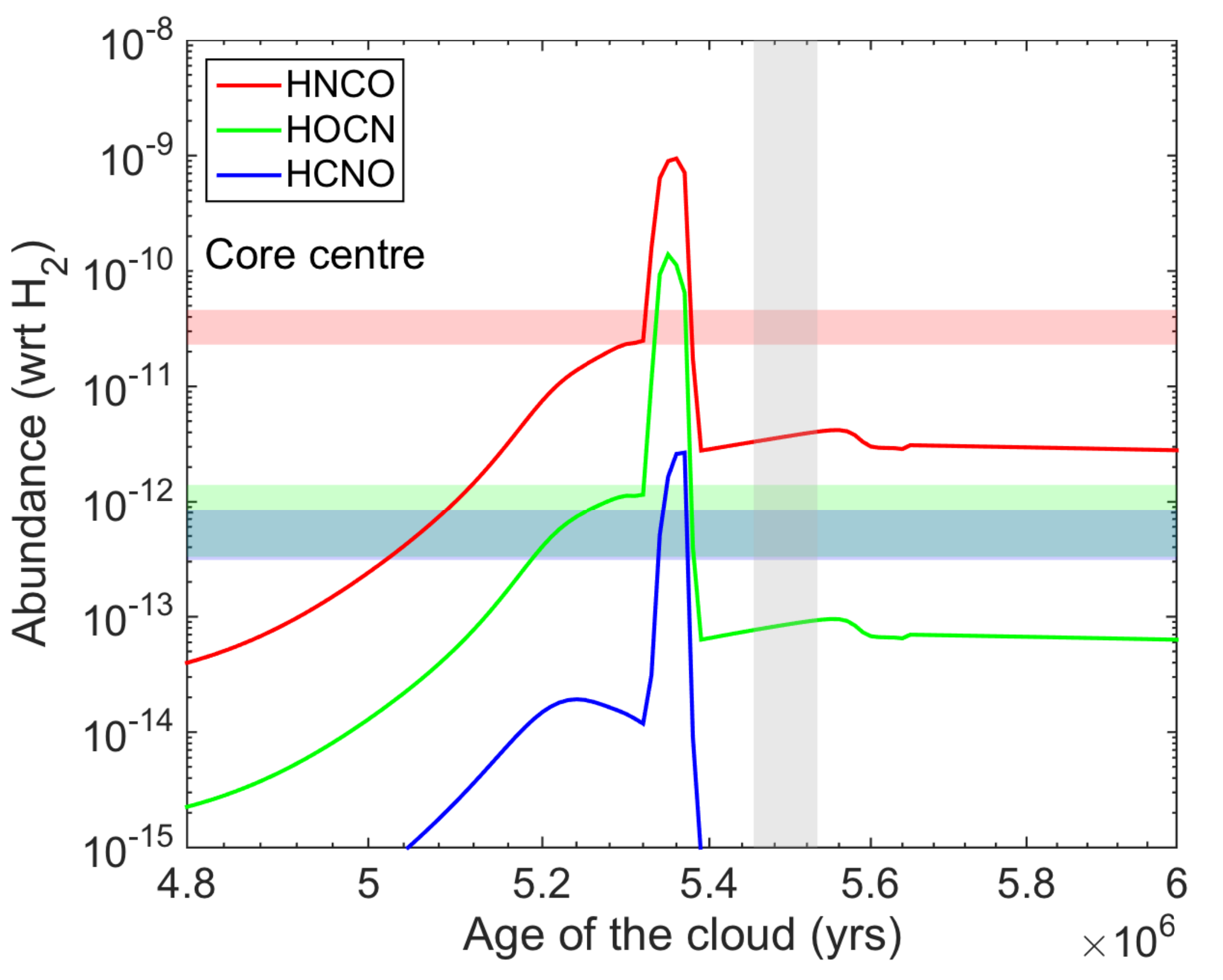}
	\includegraphics[width=0.49\hsize,clip=true,trim=0 0 0 0]{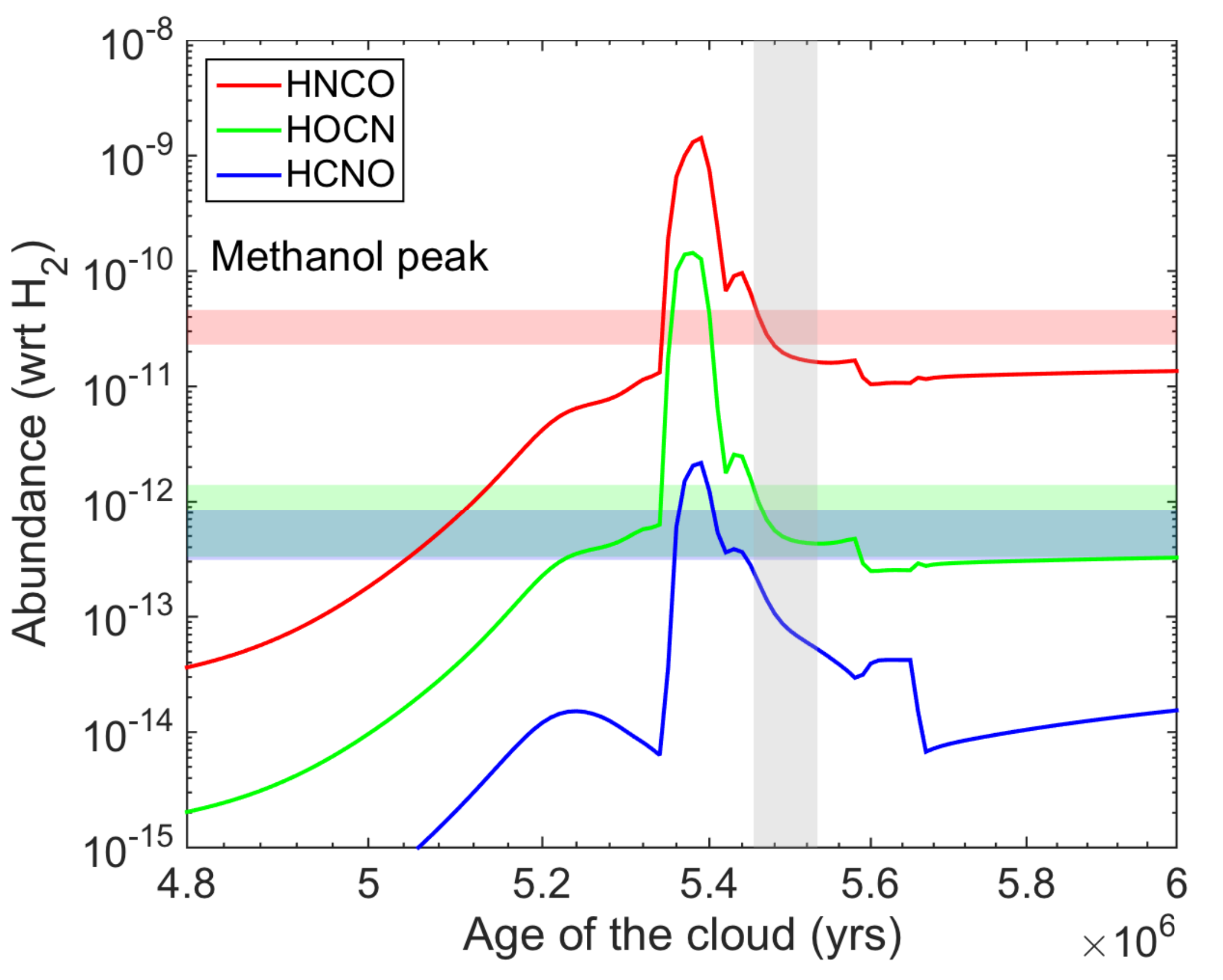}
	\caption{Final abundances as a function of time for HNCO and its isomers for the four environments. The time-scale for which we obtain the best agreement between the modelling and observations is shown in vertical grey scale. Observational constraints are shown in horizontal coloured area or in dashed lines for upper limits. For each source, the observed abundances are taken from the literature. \textit{Top left:} IRAS16293 B hot corino: \citet{martin-domenech2017}. \textit{Top right:} IRAS16293 cold envelope: the observed abundance of HNCO is calculated from the derived column density of HNCO reported by \citet[][N(HNCO)=3.4$\times$10$^{13}$\,cm$^{-2}$]{marcelino2010} and the H$_2$ column density estimated by \citet{van_dishoeck1995} for the cold envelope (N(H$_2$)=3.4$\times$10$^{13}$\,cm$^{-2}$). Analogously, the HOCN and HCNO abundances are inferred from the upper limits to the column densities of these molecules calculated by \citet{marcelino2010}. \textit{Bottom left:} L1544 core centre: \citet{marcelino2010}. \textit{Bottom right:} L1544 methanol peak: assumed the same as that of the core centre.}
	\label{HNCO_models}
\end{figure*}

The results of the best fit model to the observed abundances of HNCO and its isomers for the four different environments considered here, are shown in Figure \ref{HNCO_models}. The hot corino abundances (top left panel) of HNCO as well as the upper limits for HCNO and HOCN are reproduced within the best fit time-scale of $[2.3-4]\times$10$^4$\,yrs. The abundance of HCNO varies by several orders of magnitude in this time-scale and it even goes slightly above the upper limits (within a factor of $\sim$4) for a short period of time. Afterwards, its abundance drops quickly to values $\leq$10$^{-14}$. HOCN is well below the upper limit during this period. The sharp increase of the abundances of these molecules at $\sim$2$\times$10$^4$\,yrs is due to the increase of the temperature in the hot corino above 100\,K, which triggers the release of numerous radicals into the gas phase induced by the desorption of water.

For the cold envelope of IRAS16293, the HNCO and HOCN abundances are also correctly reproduced, during the same period of time as the hot corino. Again, HCNO goes above the upper limits (within a factor of $\sim$3.5) but this time for an extended period of time, and it does not go below the observed upper limit during our best fit time-scale.

For both regions (hot corino and cold envelope), the HNCO and HOCN observed abundances (taken from \citealp{marcelino2010} and \citealp{martin-domenech2017}) are well reproduced while HCNO seems to lay above the measured upper limits. Even though these differences are really small (less that a factor of 4 in both cases), it shows that HCNO is still slightly overestimated. Finally, the time-scale of IRAS16293 predicted by our modelling (of a few 10$^4$\,yrs), is consistent with the evolutionary stage of this object \citep{jaber_al-edhari2017}. In the hot corino, HNCO and its isomers are produced $\sim$75\% in the gas phase and $\sim$25\% on grain surfaces. The fraction formed on grains is then released into the gas phase via thermal desorption when the temperature reaches 100\,K. However, in the cold envelope, grain surface and gas-phase reactions are equally efficient. The fraction of HNCO (and of its isomers) formed on grains is non-thermally desorbed mainly via cosmic ray induced UV-photons. Chemical-reactive desorption is not efficient in the release of HNCO in the cold envelope due to the high binding energy of this molecule to the grain surface. \\

For the two positions of the L1544 pre-stellar core, the best agreement between the observations and the modelled abundances is obtained $[1-2]\times$10$^5$\,yrs after reaching the final density during the collapse. At 5.35$\times$10$^6$\,yrs the final density is reached, allowing species to be efficiently produced on grain surfaces since the high visual extinction prevents UV photons from destroying or desorbing molecules. These newly and rapidly formed species are then injected in the gas phase thanks to non-thermal desorption (such as chemical reactive desorption), explaining the sharp peak of abundances seen at this time. Later on, these species are frozen back onto the grain surface, dramatically decreasing their gas phase abundances.

Single-dish observations of HNCO and its isomers cannot disentangle whether the emission of these molecules comes from an external layer coincident with the methanol peak or from the core centre. Therefore, we used the same observed abundances for both cases (extracted from \citealp{marcelino2010}). For the methanol peak position (lower right panel), the HNCO and HOCN abundances are well fitted but HCNO is underestimated by a factor of $\sim$1.5 at least. While the HNCO and HOCN abundances remain rather constant within the best fit time-scale, HCNO is efficiently destroyed and its abundance drops by one order of magnitude in the time range of $[1-2]\times$10$^5$\,yrs after the end of the collapse.
In contrast, for the core centre position, all species present much lower abundances than the observed values. This suggests that these molecules are likely formed in an external layer and not in the core centre, as also suggested for O-bearing COMs \citep{vastel2014, jimenez-serra2016}. Moreover, recent observations performed by \citet{spezzano2017} using the IRAM 30m telescope shows emission maps of HNCO toward this source. Even though the main emission region of HNCO is not coming from the same position where methanol peaks, it clearly comes from a region $\sim$30\,arcsec ($\sim$4000\,au) away from the core centre, where the physical conditions are similar (see their Figure 1).\\

We note that the set of reactions modified to obtain a better agreement with the HCNO and HOCN observations were selected since they are the most efficient forming or destroying these two species. However, other reactions, not included yet in our network, might play an important role in the determination of the abundances of HCNO and HOCN. Indeed, \citet{eshchenko2002} and \citet{glarborg1998} have shown that the $\rm HCCO + NO \leftrightarrows HCNO + CO$ could be an important pathway to form or destroy HCNO. The recent detection of unexpected high abundances of HCCO toward cold dark clouds\citep{agundez2015} favours this idea. \citet{glarborg1998} also suggested that HCNO could be efficiently recycled to NO through the reaction with OH. The two latter reactions are not included yet in the present network since further experimental investigations of reactions involving HCNO and HOCN are needed. 

Finally, although the chemistry of HONC is included in the network, we do not analyse it here since its predicted abundances are always very low ($\lesssim10^{-15}$) for all the different models. Moreover, only upper limits are given in the literature for this isomer, showing that it is unlikely to be abundant.

\subsection{\methyl and isomers}\label{result_methyl}

As presented in Sect. \ref{network_methyl}, the \methyl gas phase network comes from \citet{halfen2015}. They proposed that reaction (\ref{methyl_grain1}) could have a reaction rate up to $\sim$10$^{-10}$\,cm$^3$\,s$^{-1}$ \citep{woodall2007}. Since we also included the isomers of \methyl in our study, as first approximation we consider that CH$_3$OCN and CH$_3$CNO follow similar formation and destruction routes to those of CH$_3$NCO (see Table \ref{table_reac_gas}).
We have varied the reaction rate of these reactions to obtain satisfactory results for the measured upper limits of CH$_3$CNO and CH$_3$OCN toward the hot corino of IRAS16293 \citep{martin-domenech2017, ligterink2017}. We have used the same reaction rates for reactions between HNCO/HOCN/HCNO/HONC and CH$_3$ to produce CH$_3$NCO/CH$_3$OCN/CH$_3$CNO/CH$_3$ONC, respectively. The final reaction rate that best fit the observations is $5\times10^{-11}$\,cm$^3$\,s$^{-1}$. We also included the following set of reactions in the gas phase:
\begin{ceqn}
	\begin{eqnarray}
		\mathrm{HNCO + CH_3} &\longrightarrow& \mathrm{CH_3OCN}\\
		\mathrm{HOCN + CH_3} &\longrightarrow& \mathrm{CH_3NCO}\label{hocn_ch3}\\
		\mathrm{HCNO + CH_3} &\longrightarrow& \mathrm{CH_3ONC}\\
		\mathrm{HONC + CH_3} &\longrightarrow& \mathrm{CH_3CNO}
	\end{eqnarray}
\end{ceqn}
These reactions yield large amounts of CH$_3$OCN, CH$_3$CNO, and CH$_3$ONC for the hot corino model and thus we had to lower their efficiencies down to $\sim$10$^{-20}$\,cm$^3$\,s$^{-1}$. Indeed, as shown in \citet{martin-domenech2017}, CH$_3$OCN and CH$_3$CNO are not detected in the hot corino of IRAS16293 with upper limits $1.75\times10^{-11}$ and $9.3\times10^{-13}$. This suggests that these reactions may not be occurring. New experimental data will be needed to test this finding. In any case, if reaction (\ref{hocn_ch3}) is set to $5\times10^{-11}$\,cm$^3$\,s$^{-1}$ (same as for $\rm HNCO + CH_3$), the final abundance of \methyl is only a factor of $\sim$\,3 off the observed abundance, which is still a satisfactory result compared to observations.\\

\begin{figure*}
	\centering
	\includegraphics[width=0.49\hsize,clip=true,trim=0 0 0 0]{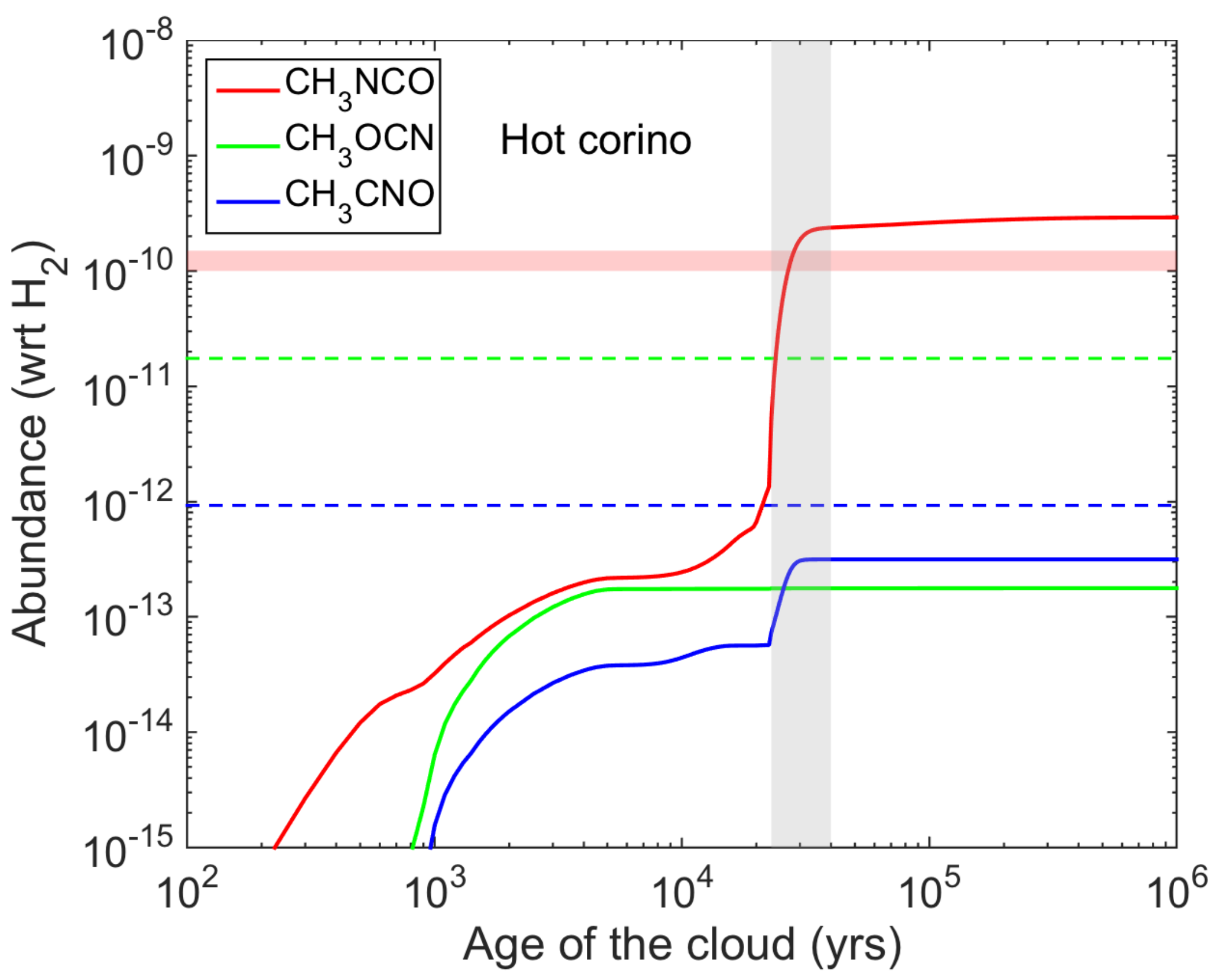}
	\includegraphics[width=0.49\hsize,clip=true,trim=0 0 0 0]{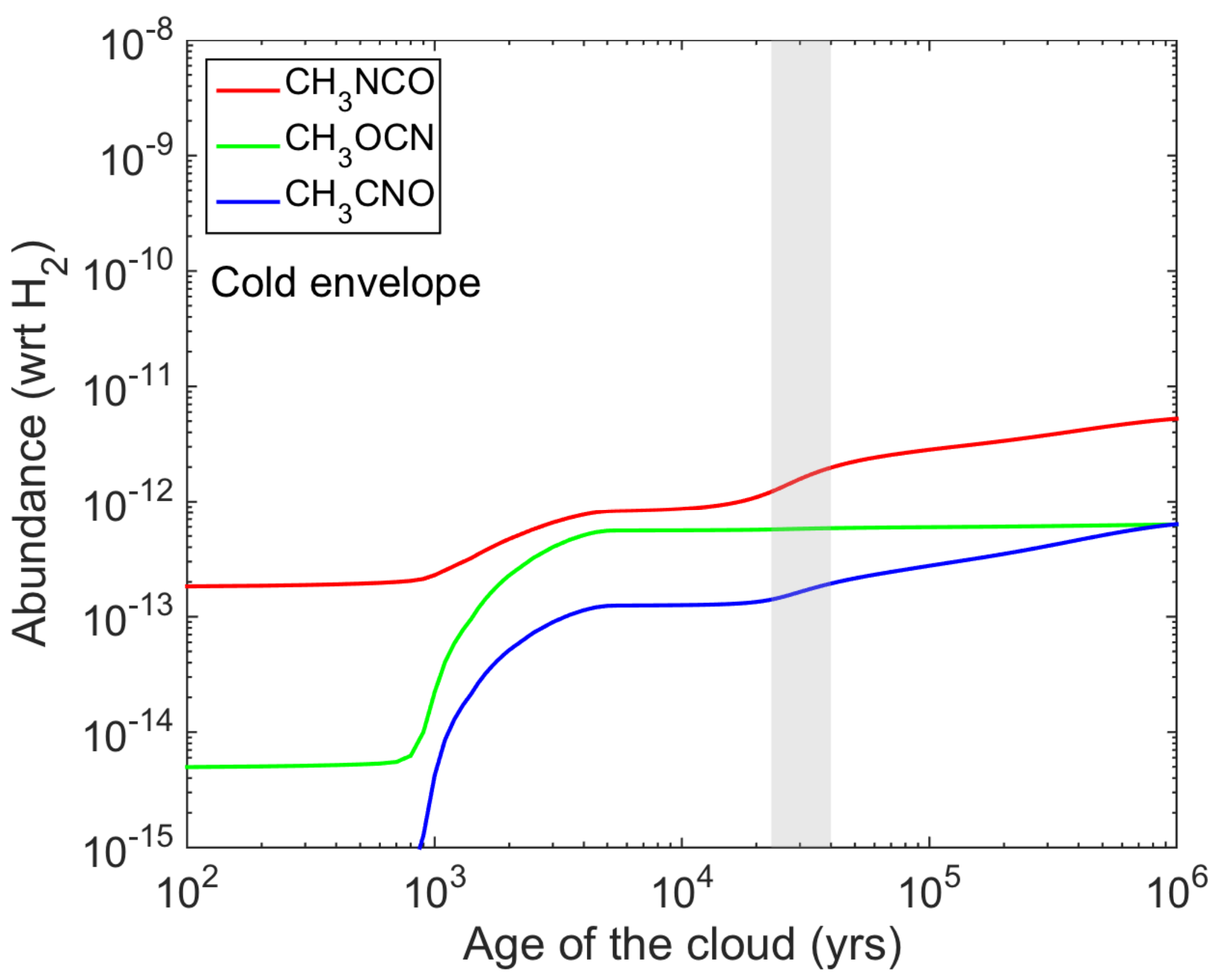}\\
	\includegraphics[width=0.49\hsize,clip=true,trim=0 0 0 0]{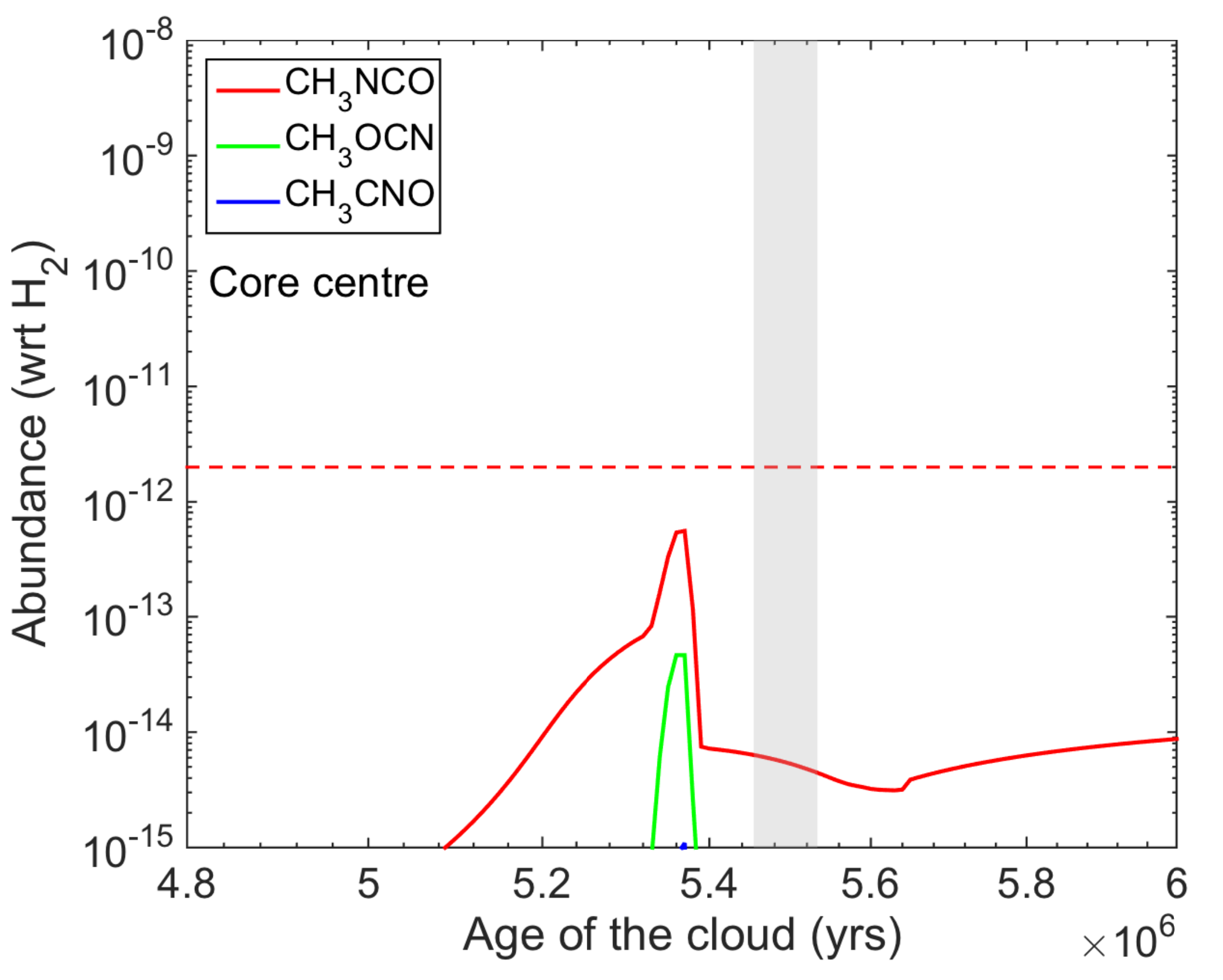}
	\includegraphics[width=0.49\hsize,clip=true,trim=0 0 0 0]{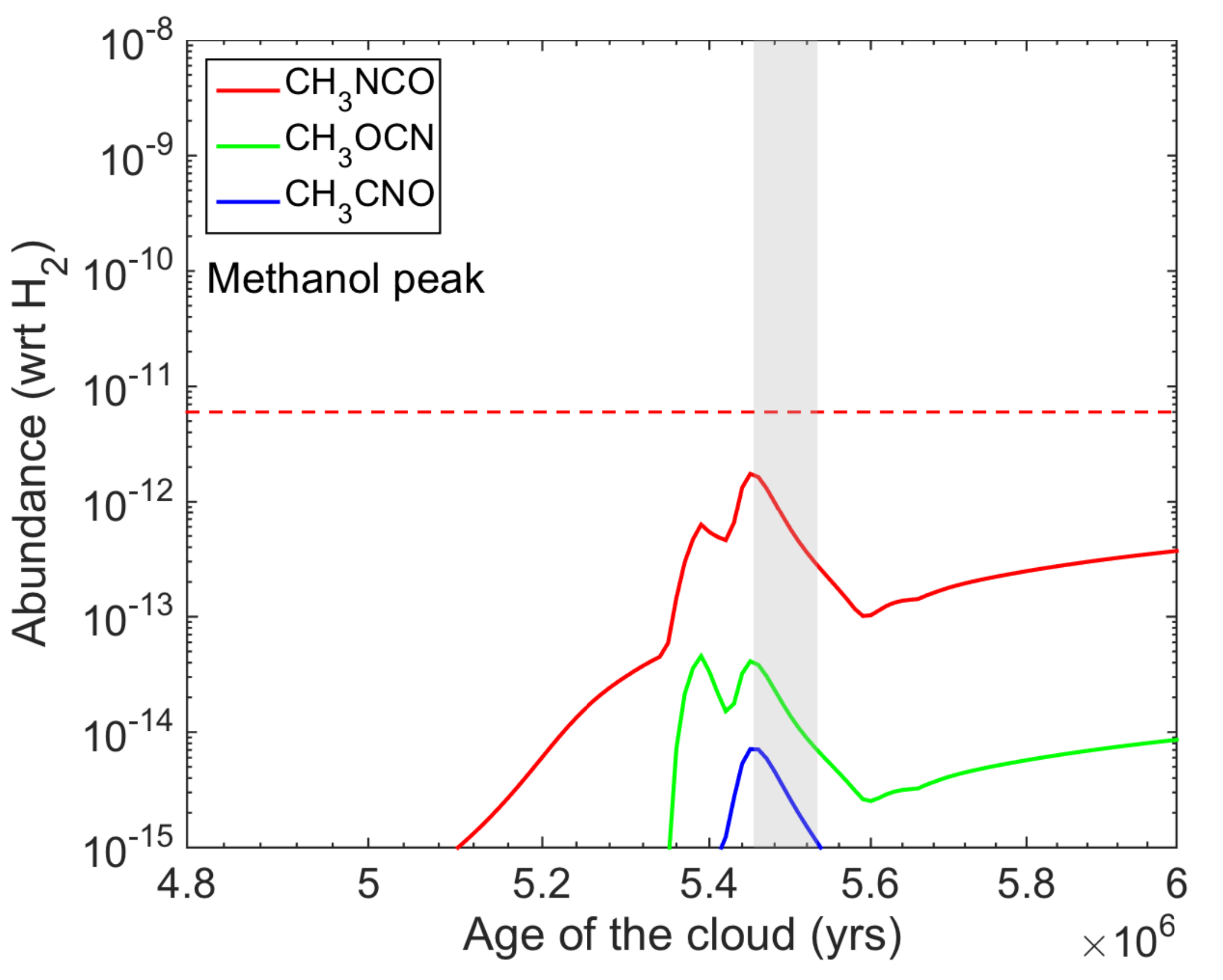}
	\caption{Same as Fig. \ref{HNCO_models} for \methyl and its isomers. \textit{Top left:} IRAS16293 B hot corino: \citet{martin-domenech2017, ligterink2017}. \textit{Top right:} IRAS16293 cold envelope: no observational constraints. \textit{Bottom left:} L1544 core centre: \methyl \citep{jimenez-serra2016}. \textit{Bottom right:} L1544 methanol peak: \methyl \citep{jimenez-serra2016}.}
	\label{methyl_models}
\end{figure*}

The results of the modelling for \methyl and its isomers, for the four different environments, are shown in Figure \ref{methyl_models}. Since \methyl has been detected recently in the ISM, only a few observational constraints exist for the different regions investigated here.
\methyl is clearly detected around the hot corino of IRAS16293 (top left panel), with an upper limit abundance of $[1.0-1.5]\times$10$^{-10}$ \citep{martin-domenech2017,ligterink2017}. In our model, the \methyl abundance sharply increases around 2$\times$10$^{4}$\,yrs. Indeed, once the temperature reaches 100\,K, species such as CH$_3$ or HNCO (and its isomers) are thermally desorbed from the grain surface, enhancing the formation of \methyl in the gas phase. From our analysis, CH$_3$NCO is produced in large quantities on the grain surface but it is then efficiently destroyed again on the surface through reaction (\ref{ch3nco_h}). This conclusion is supported by recent experimental results where methylamine is detected while performing experiments on CH$_4$:HNCO mixtures at 20\,K (N. Ligterink, private communication). Another possible destruction pathway on grain surfaces may involve the formation of N-methylformamide from successive hydrogenation of methyl isocyanate \citep{belloche2017}:
\begin{ceqn}
	\begin{eqnarray}
		\mathrm{\#CH_3NCO + \#H} &\longrightarrow& \mathrm{\#CH_3NHCO}\\
		\mathrm{\#CH_3NHCO + \#H} &\longrightarrow& \mathrm{\#CH_3NHCHO}
	\end{eqnarray}
\end{ceqn}
However, N-methylformamide is not detected as a product in the experiments of the CH$_4$:HNCO mixtures at 20\,K (N. Ligterink, private communication), suggesting that the hydrogenation of \methyl could be ineffective. We note that models including the destruction reactions of \methyl into N-methylformamide provided similar results to those considering the methylamine destruction route. In either case, the main contribution to the final gas phase abundance of \methyl comes from gas-phase reactions.

The predicted abundances of CH$_3$OCN and CH$_3$CNO are orders of magnitude lower than that of CH$_3$NCO as a result of the low abundances of HOCN and HCNO in the gas phase. The \methyl abundance and the upper limits of CH$_3$OCN and CH$_3$CNO are well reproduced by our model for the same time-scales as those inferred for HNCO and its isomers. No observational constraints exist for the cold envelope of IRAS16293 for \methyl and its isomers, but we give here the expected abundances. These species should be difficult to detect in a cold environment, as confirmed by the lack of detection of this species toward the pre-stellar cores L1544 \citep[$\leq$2-6$\times10^{-12}$;][]{jimenez-serra2016} and B1-b \citep[$\leq$2$\times10^{-12}$;][]{cernicharo2016}. Indeed, the predicted abundance of \methyl for both regions (methanol peak and core centre) is well below the upper limits estimated by \citet{jimenez-serra2016}.

As for HONC, the chemistry of CH$_3$ONC is included in the network but its predicted abundance is even lower than that of HONC ($\lesssim10^{-15}$), showing that it is unlikely to be abundant in these regions.

\subsection{\formamide}\label{result_formamide}

As presented in Section \ref{network_formamide}, the main mechanisms proposed to form formamide are hydrogenation of HNCO \citep{charnley1997,raunier2004}, the radical-radical reactions of NH$_2$ and HCO/H$_2$CO on the surface of dust grains \citep{fedoseev2016}, and the gas-phase reaction $\rm NH_2 + H_2CO \rightarrow NH_2CHO + H$ \citep{barone2015,skouteris2017}. When considering these three formation routes in our models, discrepancies (by more than a factor of 10) clearly arise between the predicted and observed abundances of formamide for the cold envelope and the L1544 models. At low temperatures ($\leq$20 K), the formation of this molecule is driven by hydrogenation and/or gas-phase formation. 

\begin{figure*}
	\centering
	\includegraphics[width=0.49\hsize,clip=true,trim=0 0 0 0]{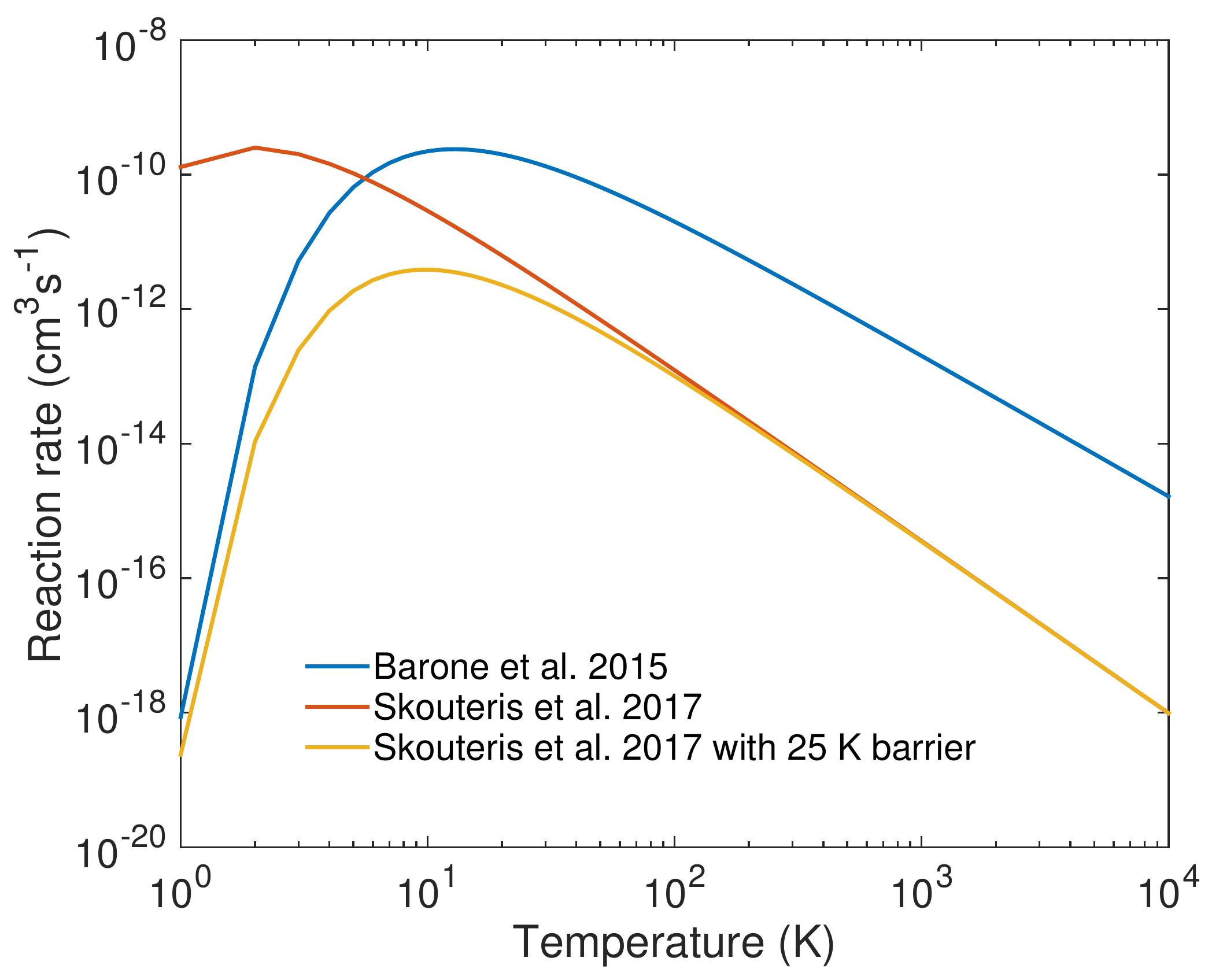}
	\includegraphics[width=0.49\hsize,clip=true,trim=0 0 0 0]{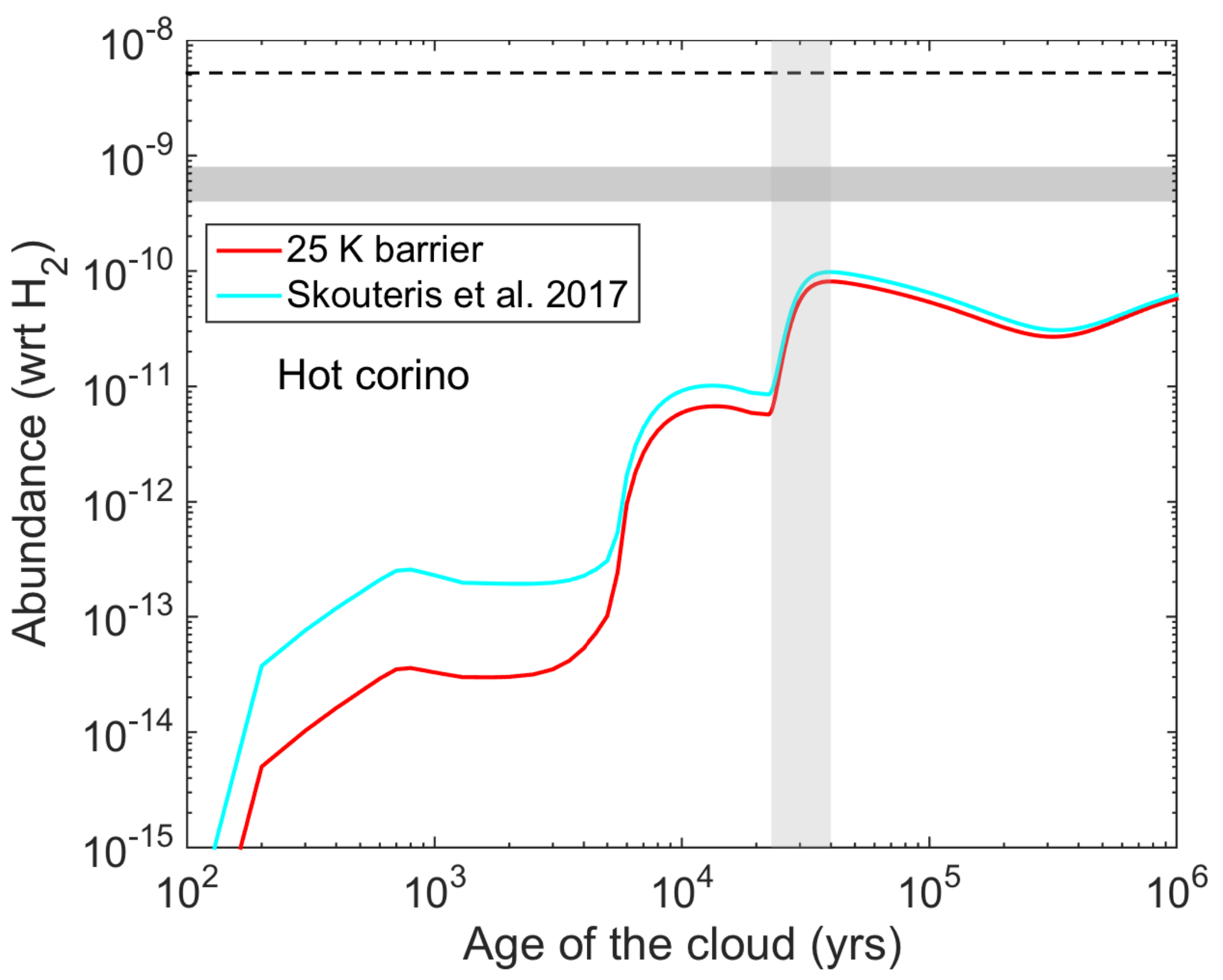}\\
	\includegraphics[width=0.49\hsize,clip=true,trim=0 0 0 0]{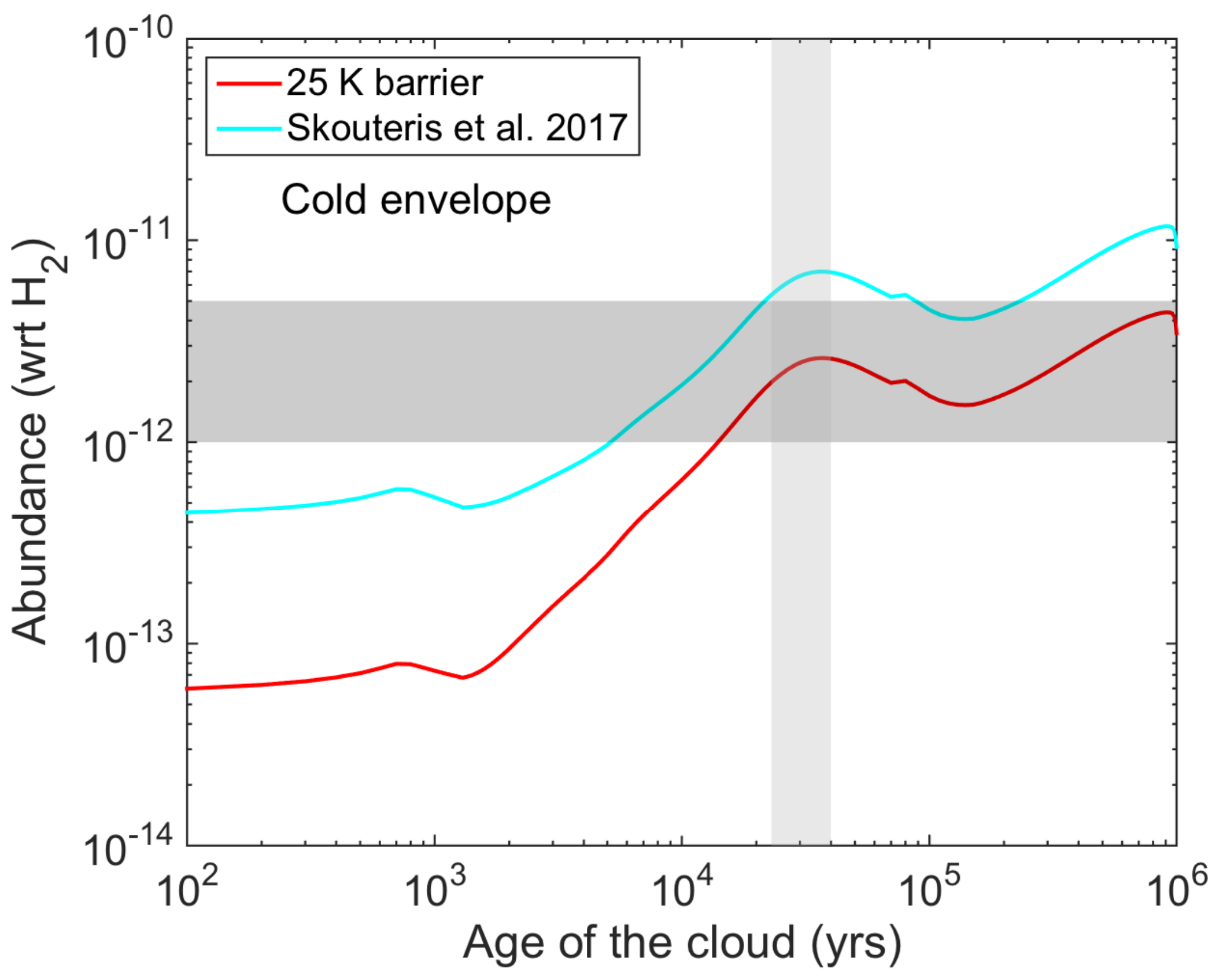}
	\includegraphics[width=0.49\hsize,clip=true,trim=0 0 0 0]{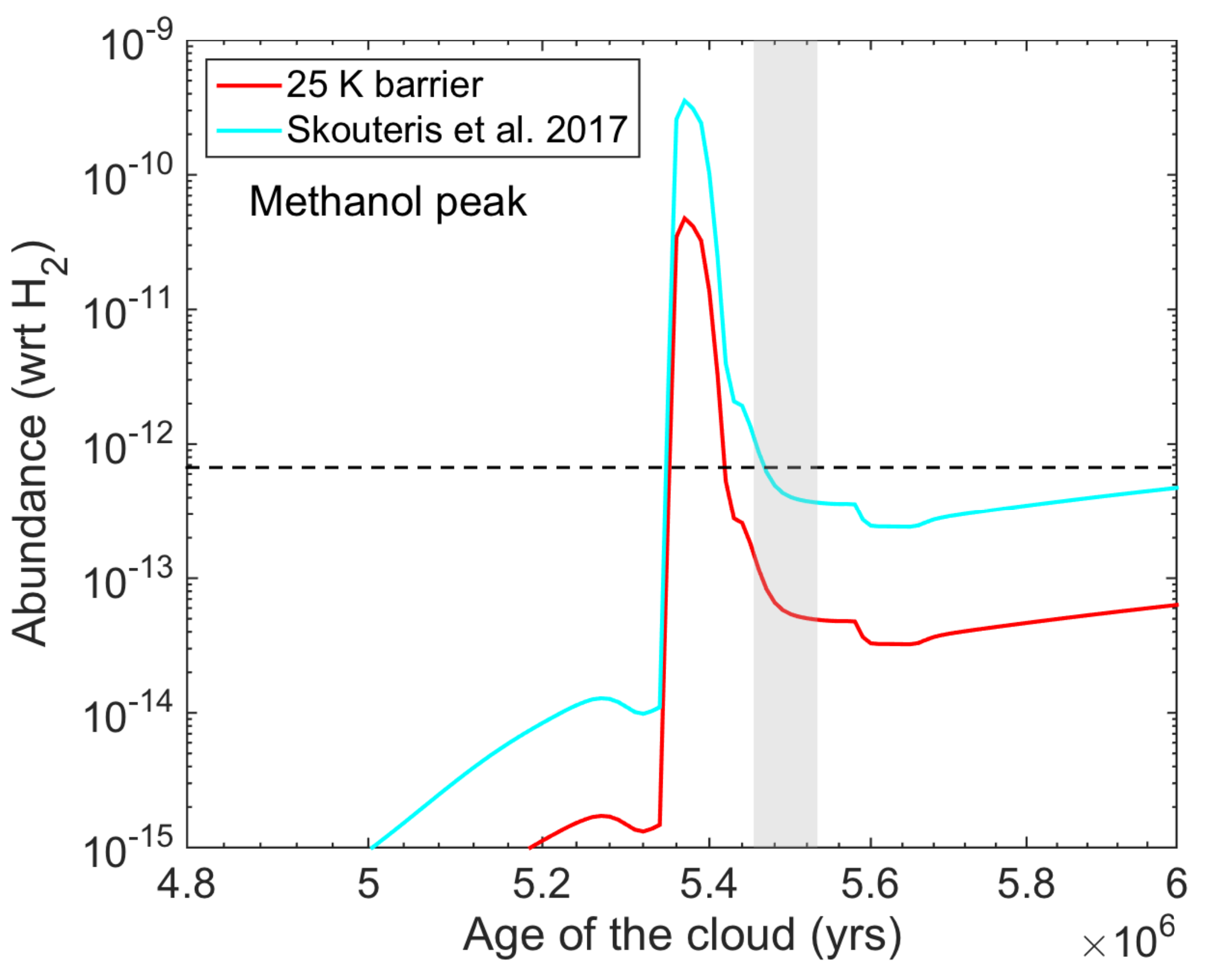}
	\caption{\textit{Top left panel:} Reaction rates of $\rm NH_2 + H_2CO \longrightarrow NH_2CHO$ as a function of temperature for several rates taken from the literature (in blue and red) and our study (in yellow). \textit{Other panels:} Final abundance of \formamide as a function of time for three different environments: hot corino IRAS16293 B (top right panel), cold envelope of IRAS16293 (bottom left panel), and methanol peak of L1544 (bottom right panel). The two different colours represent the rate used in our study for this reaction (in red) and the rate described in \citet[][in blue]{skouteris2017}. The time-scale for which we obtain the best agreement between the modelling and observations is shown in vertical grey scale. Observational constraints are shown in horizontal coloured area or in dashed lines for upper limits.}
	\label{cmp_barone_skouteris}
\end{figure*}

In order to obtain a better match to the observations, we investigated the possibility that the rate of the gas-phase reaction $\rm NH_2 + H_2CO \rightarrow NH_2CHO + H$ has an energy barrier slightly higher than 4.88\,K as recently estimated by \citet{skouteris2017}. In Fig. \ref{cmp_barone_skouteris} (top left panel), we compare the different reaction rates calculated by \citet{barone2015} and \citet{skouteris2017} (see red and blue lines). In this Figure (top right panel and lower panels), we also present the predicted abundances of formamide for the hot corino and cold envelope cases of IRAS16293, and for the methanol peak of L1544, considering only the production of this molecule via gas-phase reactions. From Fig. \ref{cmp_barone_skouteris}, the \citet{barone2015} rate is at least 2 orders of magnitude higher than the \citet{skouteris2017} rates for temperatures $\geq$10$\,$K, and therefore it largely overproduces formamide in all sources considered in this study (e.g. by a factor of 100 for L1544). The new \citet{skouteris2017} rate provides a better agreement with observations although a small overproduction (by a factor of $\sim$2) exists for the model of the cold envelope of IRAS16293 (see light blue line in the left lower panel of Fig. \ref{cmp_barone_skouteris}). In any case, when hydrogenation and radical-radical surface reactions are considered, this overproduction is even clearer not only for the model of the IRAS16293 cold envelope, but also for the L1544 pre-stellar core (by factors >10 for both sources). Therefore, we changed the small barrier of 4.88\,K calculated by \citet{skouteris2017} to a higher value of $25$\,K closer to the one found by \citet{barone2015} of 26.9\,K. The resulting reaction rate, which is a combination of both the \citet{barone2015} and \citet{skouteris2017} rates (see yellow curve in the top left panel of Fig. \ref{cmp_barone_skouteris}), is smaller at 10-20\,K and hence, the predicted abundance of formamide in the cold envelope of IRAS16293 is better reproduced. At higher temperatures ($\gtrsim$70\,K), our proposed reaction rate is very close to the one from \citet{skouteris2017} (see top left panel of Fig. \ref{cmp_barone_skouteris}), so that the inclusion of the 25\,K activation barrier does not affect the predicted abundances of formamide at temperature >70\,K, as shown for the hot corino model (see top right panel in Fig. \ref{cmp_barone_skouteris}). In the models below, we therefore assume an activation barrier of 25$\,$K for the gas-phase formation route of formamide $\rm NH_2 + H_2CO \rightarrow NH_2CHO + H$.  

\begin{figure*}
	\centering
	\includegraphics[width=0.49\hsize,clip=true,trim=0 0 0 0]{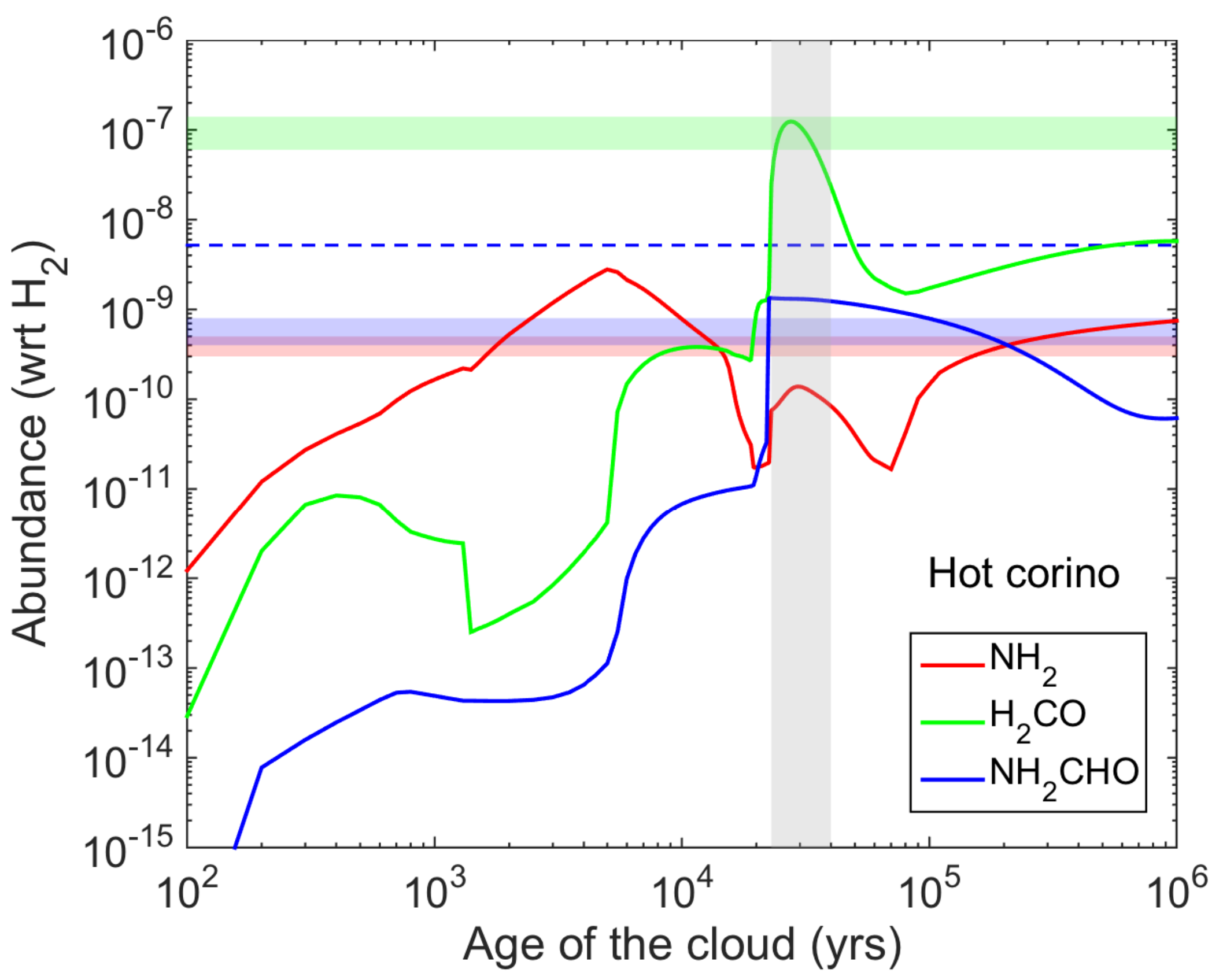}
	\includegraphics[width=0.49\hsize,clip=true,trim=0 0 0 0]{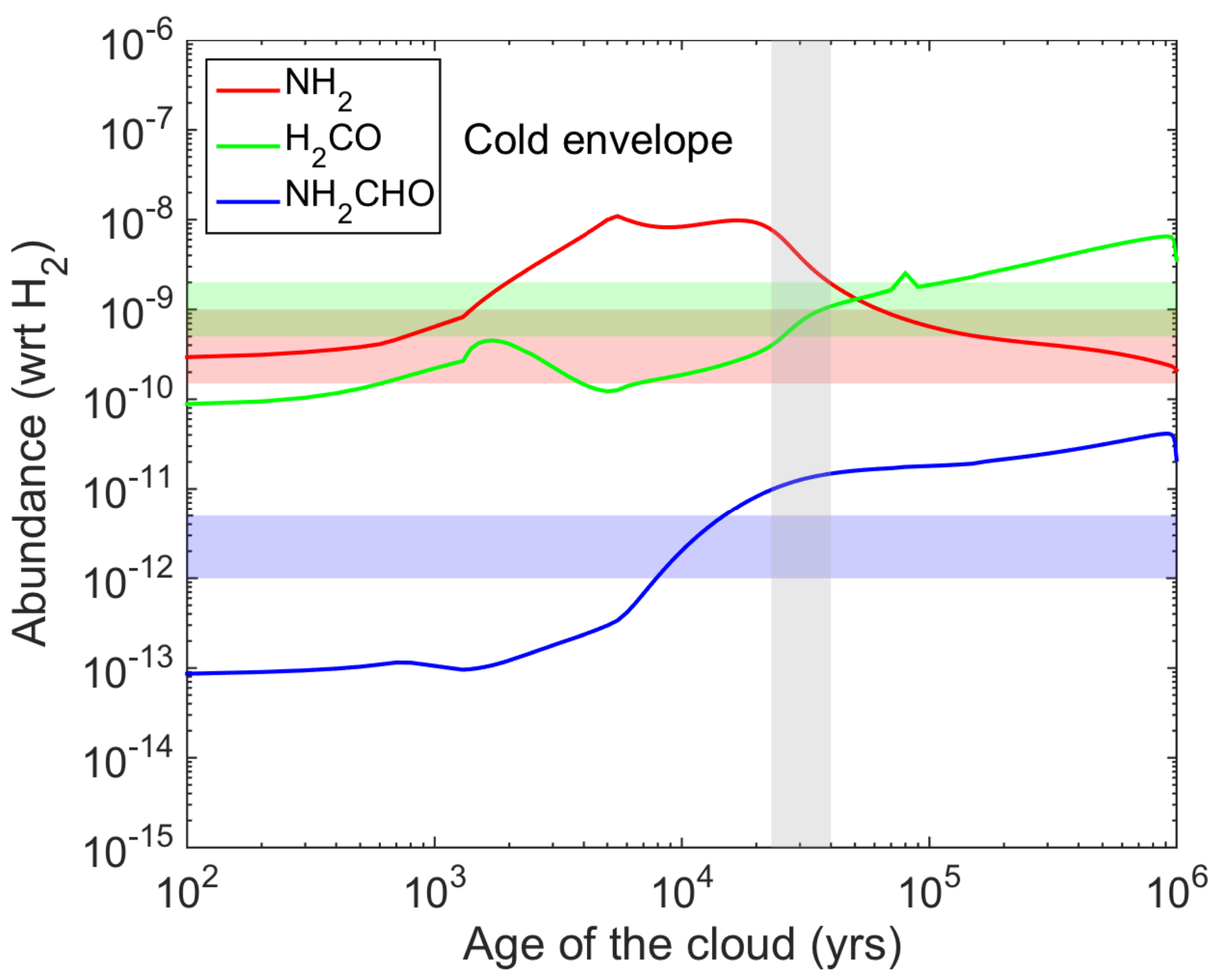}\\
	\includegraphics[width=0.49\hsize,clip=true,trim=0 0 0 0]{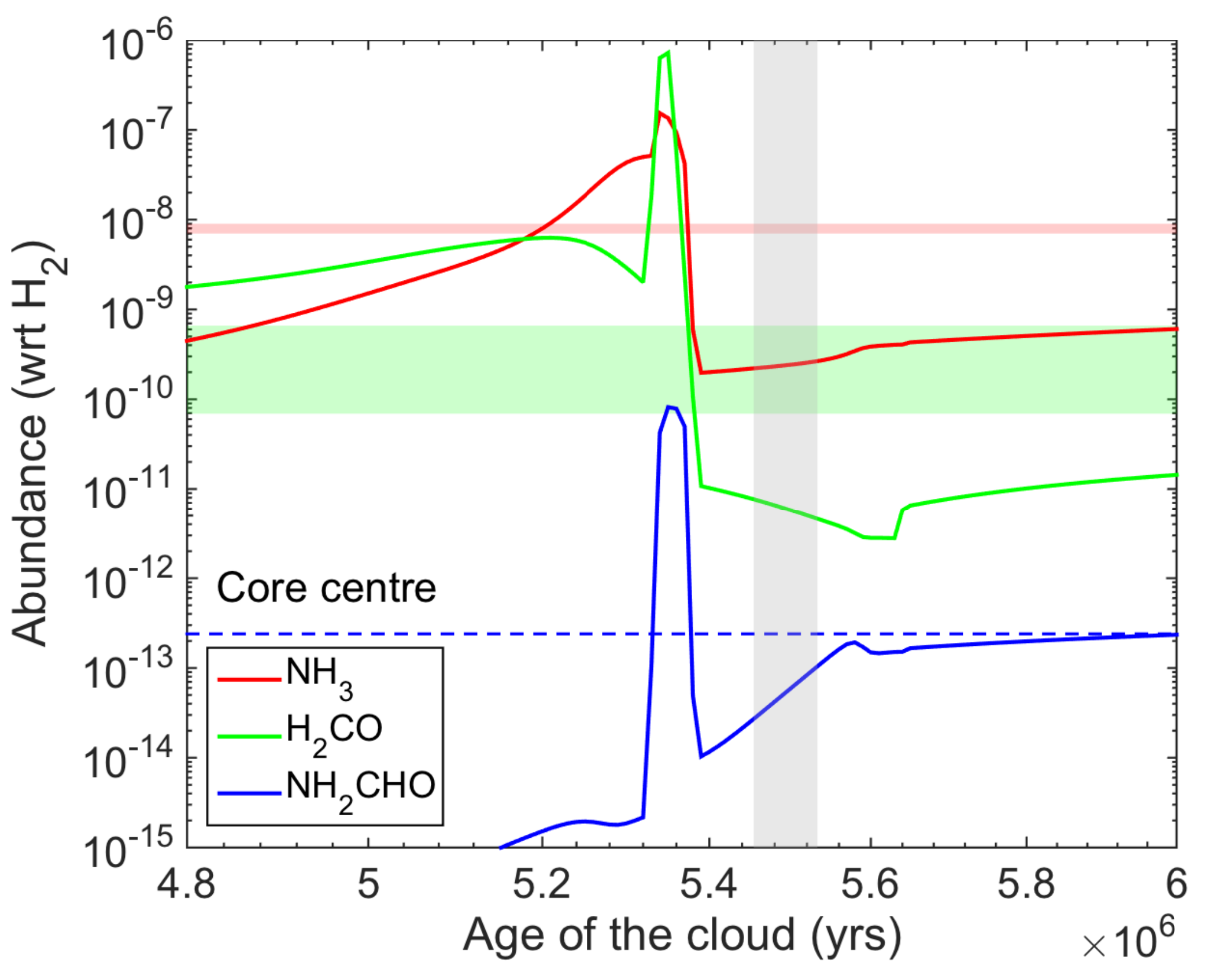}
	\includegraphics[width=0.49\hsize,clip=true,trim=0 0 0 0]{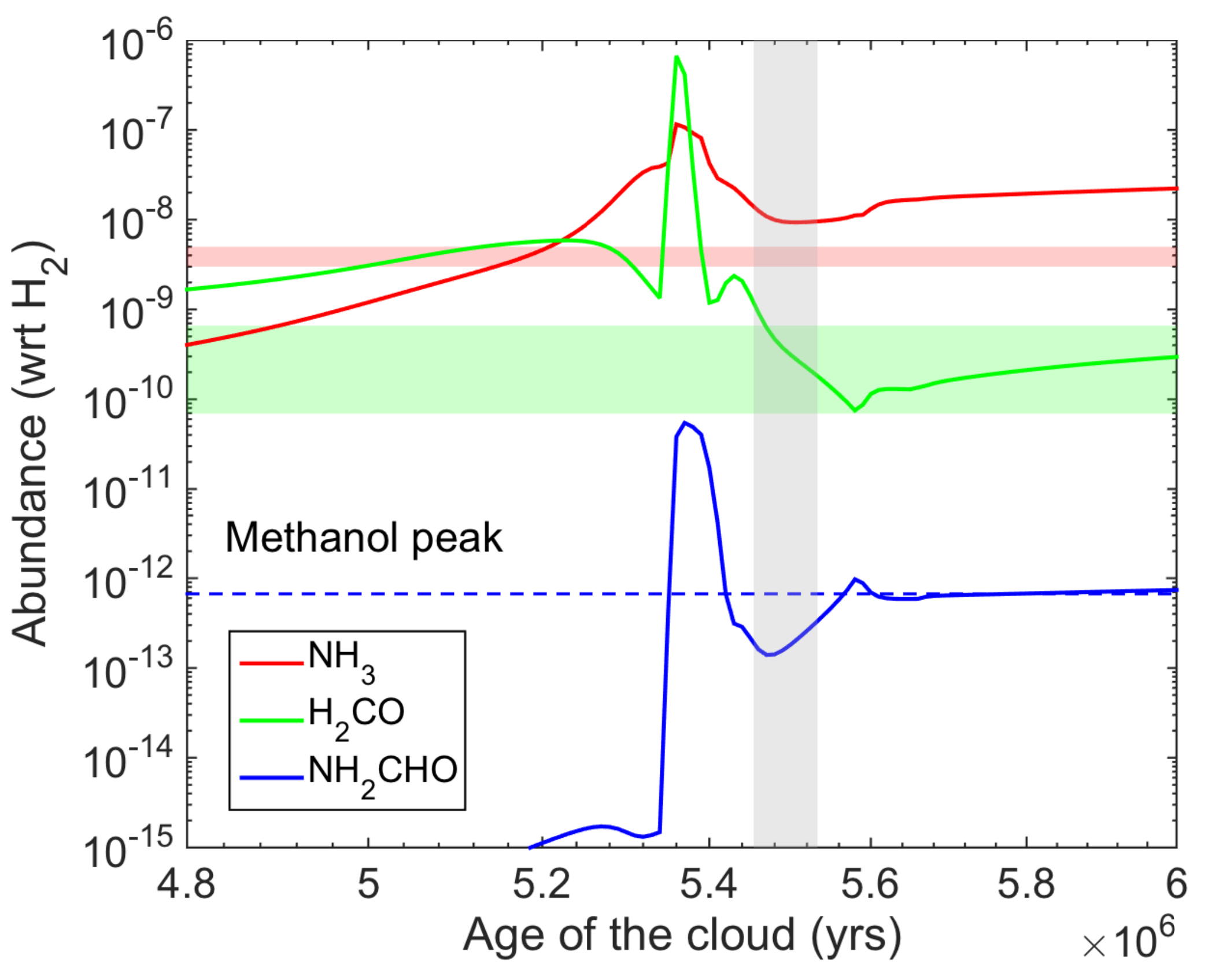}
	\caption{Same as Fig. \ref{HNCO_models} for \formamide and two parent species. \textit{Top left:} IRAS16293 B hot corino: NH$_2$ \citep{hily-blant2010}, H$_2$CO \citep{ceccarelli2000}, and \formamide \citep{lopez-sepulcre2015}. \textit{Top right:} IRAS16293 cold envelope: same references as for the hot corino. \textit{Bottom left:} L1544 core centre: NH$_3$ \citep{crapsi2007}, H$_2$CO \citep{bacmann2003}, and \formamide \citep{jimenez-serra2016}. \textit{Bottom right:} L1544 methanol peak: NH$_3$ \citep{crapsi2007}, H$_2$CO: assumed the same as that of the core centre; \formamide \citep{jimenez-serra2016}.}
	\label{formamide_models}
\end{figure*}

The results of the modelling for NH$_2$CHO, for the four different environments considered in this work, are shown in Figure \ref{formamide_models}. This figure presents the abundances of \formamide alongside those of NH$_2$ and H$_2$CO, two important parent species of formamide. For the hot corino of IRAS16293, the observed abundance of formamide was taken from  \citet[][note that this value should be considered as an upper limit]{coutens2016} and from the inferred value of \citet{lopez-sepulcre2015}. In both cases, our predicted abundance agrees well (within a factor of 1.7) with the observed values. The modelled abundance of H$_2$CO also agrees well while the NH$_2$ abundance is consistent within a factor $\sim$2.2 with respect to observations. The large enhancement of these species at $\sim$2$\times$10$^4$ yrs is again due to the thermal desorption of the ices once the temperature in the hot corino reaches 100 K. Indeed, the formamide abundance greatly increases when H$_2$CO -- a key parent species of formamide in the gas phase -- also increases. Moreover, at this temperature, all the formamide contained on the grain surface and formed at lower temperatures through radical-radical and hydrogenation reactions (with T$\gtrsim$40\,K), is also released into the gas phase.

The abundance of formamide in the cold envelope of IRAS16293 agrees well (within a factor of 3) with the observed value derived by \citet[][]{lopez-sepulcre2015}. The abundance of formamide is enhanced around $\sim$$\,$10$^4\,$yrs because of the higher temperature (20\,K), which favours the hydrogenation of HNCO. For NH$_2$ and H$_2$CO, we use the observed abundances presented in \citet{hily-blant2010} and \citet{ceccarelli2000}. While the predicted abundance of H$_2$CO matches the observations, the NH$_2$ abundance is overestimated by factors 2-8. As we discuss in Section \ref{formation_formamide}, better agreement between the predicted and observed abundances of formamide is achieved for the cold envelope of IRAS16293 when hydrogenation reactions are switched off in our models. 

For the cold core L1544, NH$_2$ abundance has not been observationally constrained and therefore, we use the chemically related species NH$_3$ to test the chemical network of NH$_2$. As for HNCO and its isomers, the H$_2$CO single-dish observations cannot disentangle whether their emission arises from the core centre or from an external layer coincident with the methanol peak position and hence, we use the same observed abundances for the two positions. By comparing the results shown in the bottom left and right panels of Fig. \ref{formamide_models}, we conclude that the emission region of H$_2$CO has to be located in an external layer of the core, since the abundances predicted for the methanol peak position reproduce the observed values. Moreover, the H$_2$CO abundance for the core centre is much lower as a result of the severe freeze out, showing that this species is not expected to be abundant in the gas phase toward this region. For NH$_3$, the predicted abundance of the methanol peak is only overestimated by a factor of 2 with respect to the value obtained by the interferometric observations of \citet{crapsi2007}. However, for the core centre, the modelled abundance is much lower than the observed value. Recently, \citet{caselli2017} also found that their chemical model predicts a much lower abundance in the central part of L1544 and they concluded that several factors might play an important role in this discrepancy, such as the underestimation of the production of gas phase NH$_3$. The observed upper limits of \formamide in L1544 \citep[$\leq$2.4--6.7$\times$10$^{-13}$;][]{jimenez-serra2016} agree well with our predictions for the estimated age of the core. As for HNCO and its isomers, the same effect is seen between [5.3--5.4]$\times$10$^6$\,yrs when the final density in the collapse is reached: the abundances of NH$_3$, H$_2$CO, and NH$_2$CHO reach their peak values to then decrease later on as a consequence of severe freeze-out onto dust grains (see Section \ref{result_hnco} for a more detailed explanation).\\

\begin{figure}
	\centering
	\includegraphics[width=0.98\hsize,clip=true,trim=0 0 0 0]{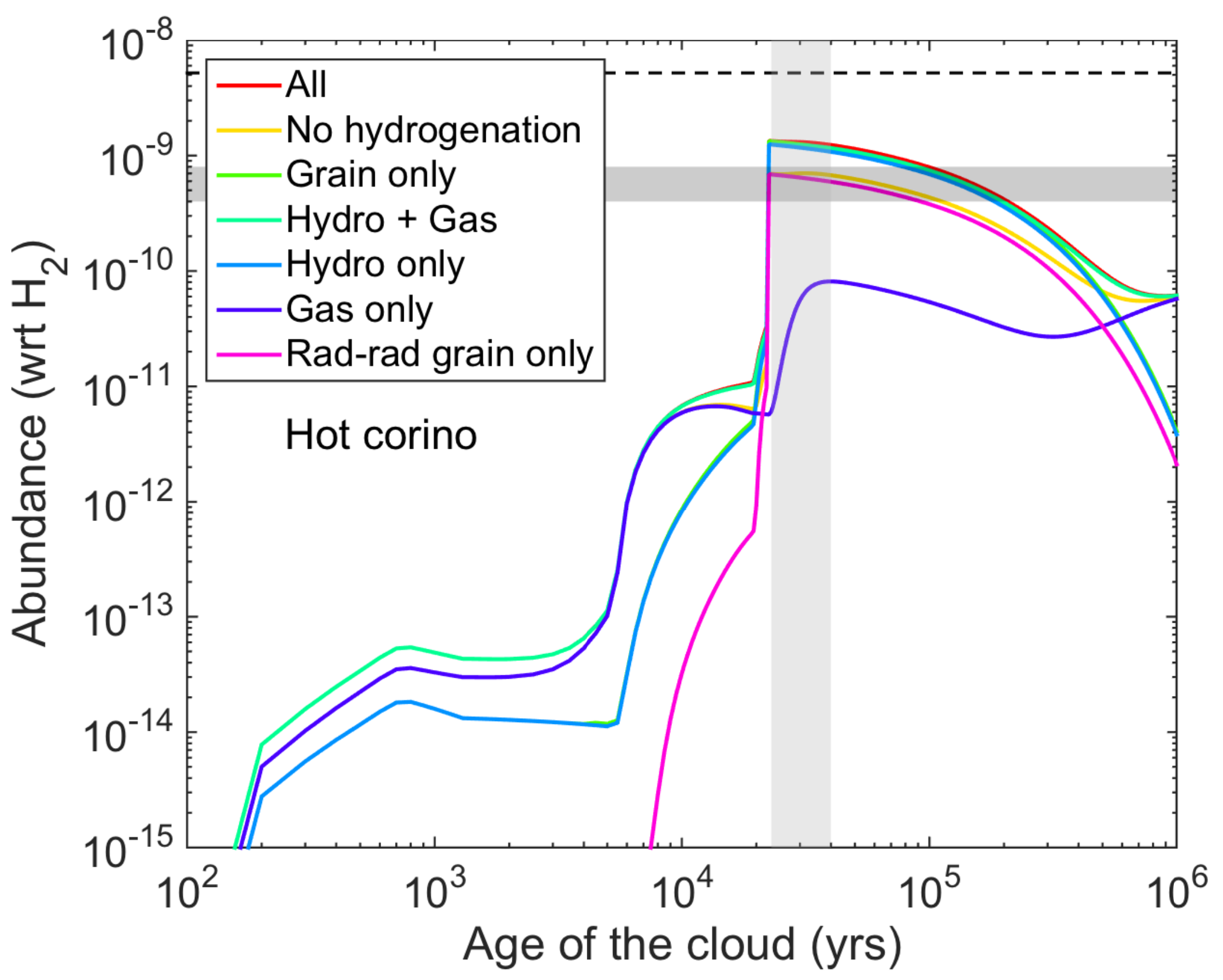}\\
	\includegraphics[width=0.98\hsize,clip=true,trim=0 0 0 0]{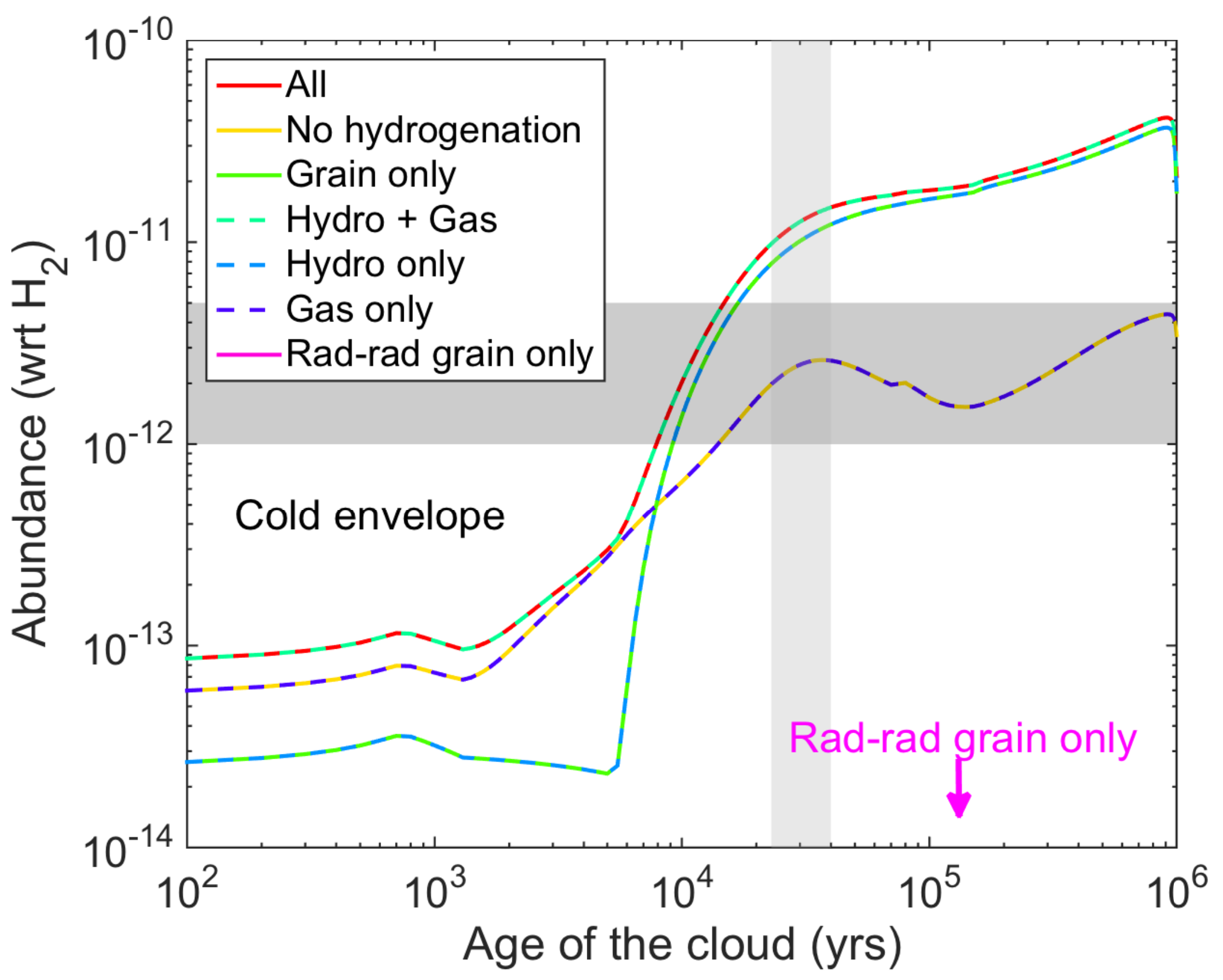}\\
	\includegraphics[width=0.98\hsize,clip=true,trim=0 0 0 0]{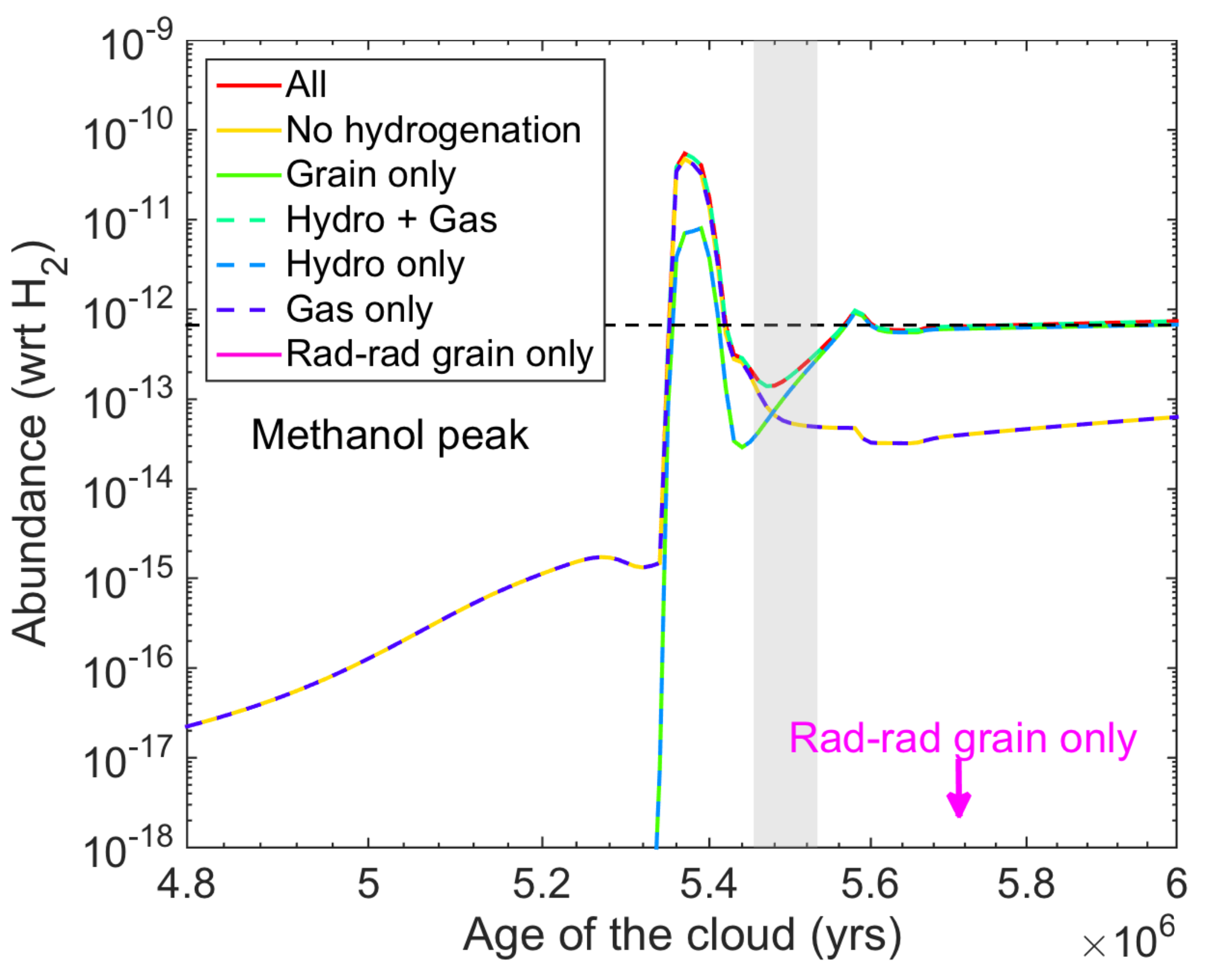}
	\caption{Final abundance of \formamide as a function of time for three different environments: hot corino IRAS16293 B (top panel), cold envelope of IRAS16293 (middle panel), and methanol peak of L1544 (bottom panel). Each colour represents a different chemical network for the formation of formamide. The time-scale for which we obtain the best agreement between the modelling and observations is shown in vertical grey scale. Observational constraints are shown in horizontal coloured area or in dashed lines for upper limits.}
	\label{compare_formamide}
\end{figure}

\section{The formation of formamide across multiple environments}\label{formation_formamide}

\begin{figure*}
	\centering
	\includegraphics[width=0.49\hsize,clip=true,trim=0 0 0 0]{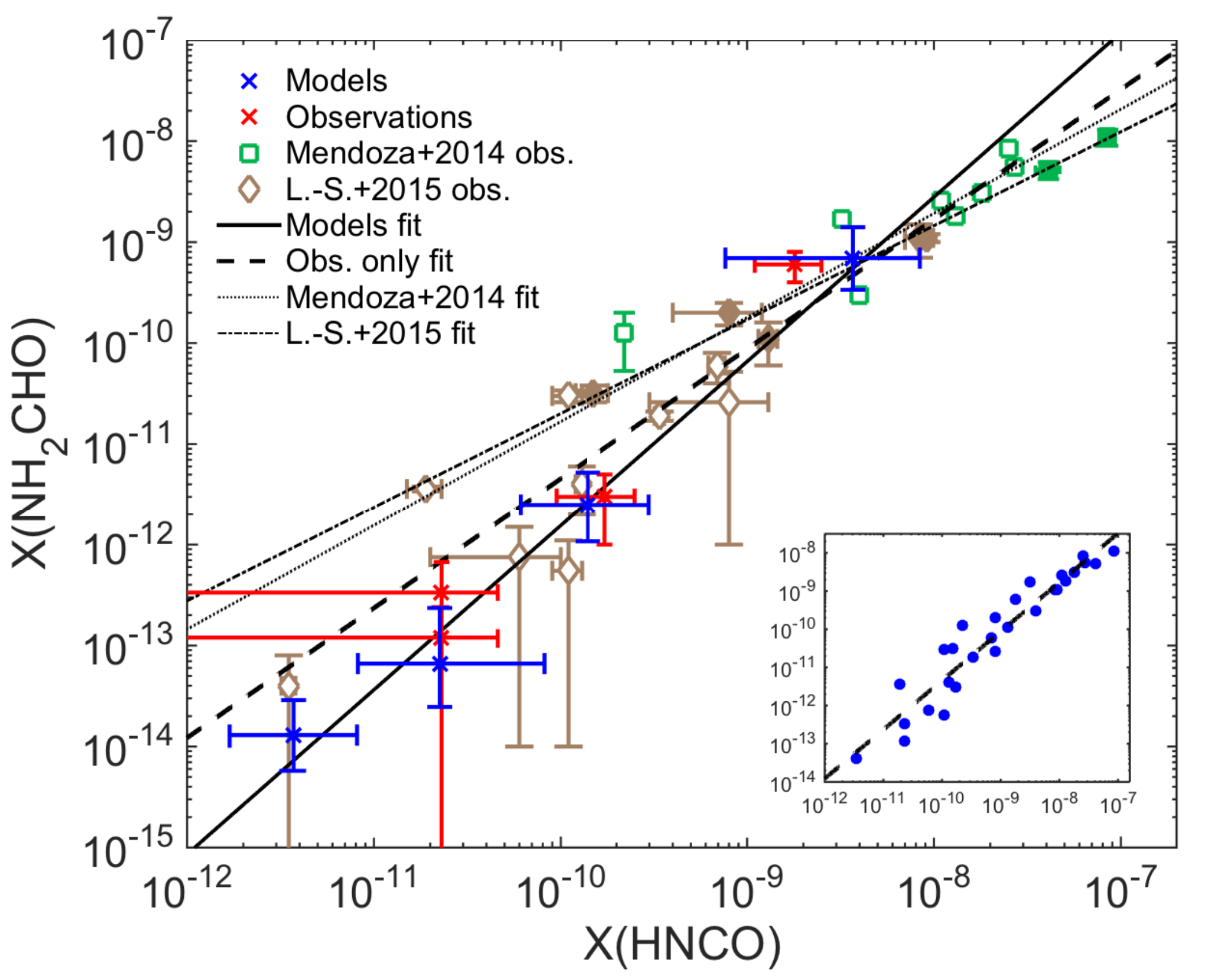}
	\includegraphics[width=0.49\hsize,clip=true,trim=0 0 0 0]{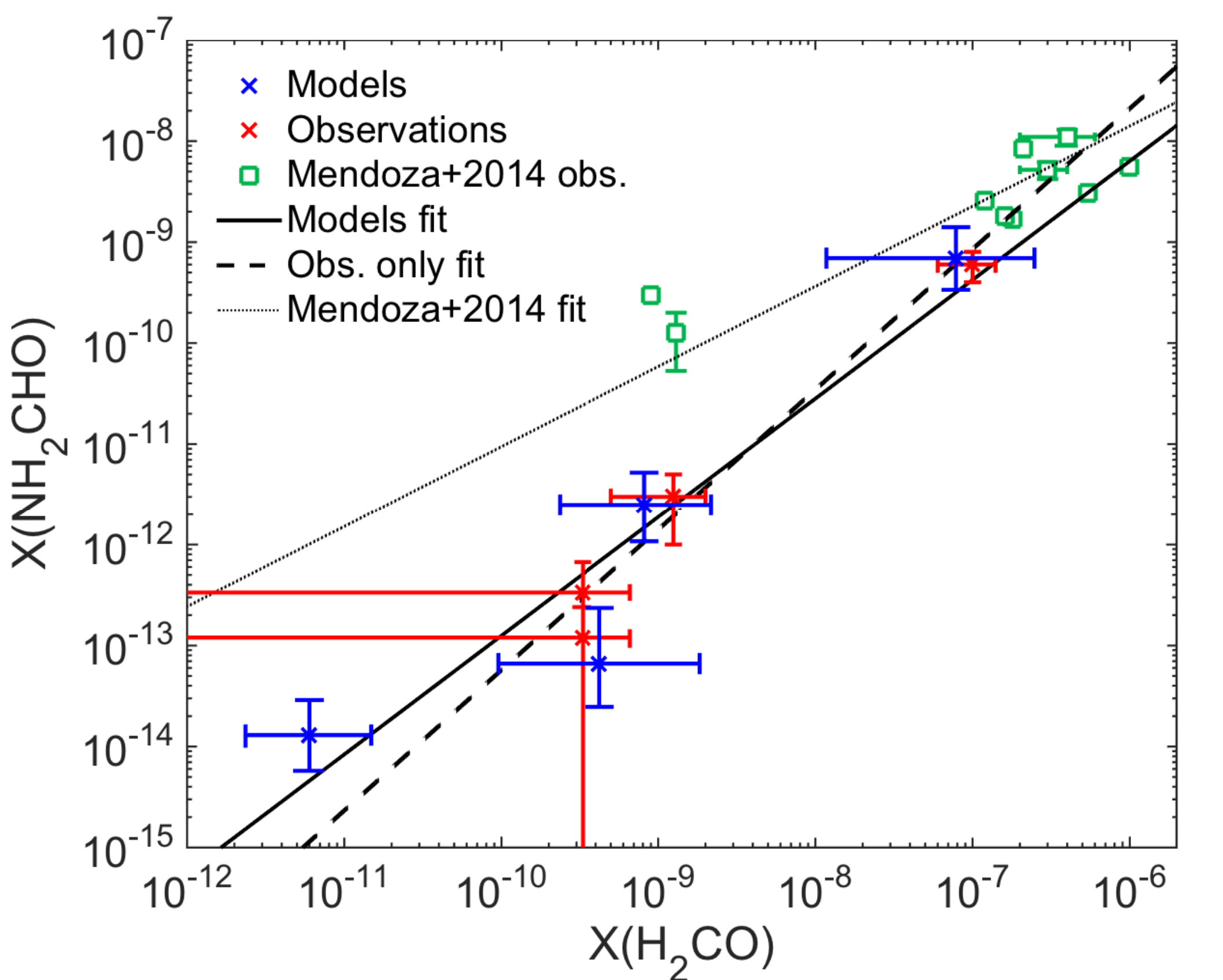}
	\caption{\textit{Left panel:} Modelled (in blue) and observed (in red) \formamide abundance as a function of HNCO for the four different environments selected in our study. Green rectangular and brown diamond points represents the value taken from \citet{mendoza2014} and \citet{lopez-sepulcre2015}, respectively. The black line shows a fitted power-law profile to the modelled abundances. The dotted and dash-dotted lines are previous fits obtained by \citet{mendoza2014} and \citet{lopez-sepulcre2015}, respectively. The black dashed line represents a fitted power-law profile to all the observations (red, green and brown points). Filled points are the one used to determine the \citet{lopez-sepulcre2015} fit. The small insert on the bottom right shows a better representation of the distribution of the observed points around the fitted profile (dashed line). \textit{Right panel:} Same as left panel with H$_2$CO instead of HNCO.}
	\label{h2co/hnco_vs_nh2cho}
\end{figure*}

The origin of formamide is a subject of strong debate and, as presented in Section \ref{network_formamide}, three main processes have been proposed for the formation of this molecule based on laboratory experiments, theoretical calculations and observations: i) the hydrogenation of HNCO on grain surfaces; ii) radical-radical formation on grains; iii) or radical-radical association in the gas phase. In the following, we investigate which of these processes are strictly needed to reproduce the observations of this molecule across multiple environments, and which processes dominate depending on the physical conditions of the source.

In Fig. \ref{compare_formamide}, we show the modelled abundance of \formamide using different combinations of these processes for the hot corino (upper panel), cold envelope of IRAS16293, and for the methanol peak of the L1544 pre-stellar core. From these figures, one notes that if we switch off one or several of these processes, the predicted abundance of formamide can vary up to an order of magnitude. For the hot corino case (Fig. \ref{compare_formamide}, top panel), the gas-phase only network does not produce enough formamide in the gas phase, indicating that grain-surface reactions may be required to match the observations. On the contrary, the network considering only grain-surface reactions gives closer values to the observed abundance, with the radical-radical network alone providing the best match (see purple line in top panel of Fig. \ref{compare_formamide}). 

For the cold envelope of IRAS16293, an interesting behaviour is found with 3 distinct group of reactions (see middle panel of Fig. \ref{compare_formamide}). Due to the low temperatures in the envelope ($\leq$20 K), the radical-radical only network predicts very low abundances of formamide (see purple arrow) because NH$_2$ and H$_2$CO are not mobile on the grain surface. Therefore, the chemistry of formamide in this region is driven either by hydrogenation or gas-phase reactions. Three options are then considered: hydrogenation only, gas phase only, and the combination of the two. The hydrogenation only network gives a higher abundance for formamide than the gas-phase one (by a factor of 3), with the gas-phase only network perfectly matching the observations. 

A similar behaviour is found for the methanol peak position of L1544 where a very low formamide abundance is expected (see lower panel in Fig. \ref{compare_formamide}). Again, radical-radical reactions are ineffective and the chemistry is only driven either by hydrogenation or gas-phase reactions. In this case, however, it is not possible to constrain which of the two processes govern the formation of formamide since both predict abundances below the measured upper limits. 

In summary, this experiment thus favours a combination of gas-phase reactions and radical-radical reactions on grain surfaces to explain the observed abundances of formamide across multiple astrophysical environments. This result is in agreement with the laboratory experiments of \citet{noble2015} and \citet{fedoseev2015}, who showed that the intermediate product $\#$H$_2$NCO formed after the hydrogenation of $\#$HNCO on grain surfaces, does not yield $\#$NH$_2$CHO but $\#$HNCO after another successive hydrogenation.

\section{The origin of the observed correlation between HNCO and NH$_2$CHO}

As discussed in Section \ref{formation_formamide}, hydrogenation reactions on the surface of dust grains are likely not responsible for the formation of formamide in the ISM. This mechanism was proposed to explain the tight correlation observed between HNCO and NH$_2$CHO in star-forming regions and reported by \citet{mendoza2014} and \citet{lopez-sepulcre2015}. In this Section, we analyse whether this correlation is due to other factors associated with the chemistry of HNCO and NH$_2$CHO.

In Figure \ref{h2co/hnco_vs_nh2cho} we show the \formamide abundance as a function of the HNCO (left panel) and H$_2$CO (right panel) abundances predicted by our model for the four different environments selected in our study. Similar plots have been presented by \citet{mendoza2014} (green rectangular points associated to the dotted line) and \citet{lopez-sepulcre2015} (brown diamond points associated to the dash-dotted line) but mainly for sources with higher abundances of formamide and isocyanic acid. The black dashed and full lines are, respectively, a power-law fit to the observed and to the modelled abundance values. In our study we derive different trends than the one derived by \citet{mendoza2014} and \citet{lopez-sepulcre2015} because we are considering the observed abundance values derived in cold sources for which the abundances of \formamide, HNCO, and H$_2$CO are lower than in other type of source (e.g. hot corino or shock regions). Indeed, even though the study of \citet{lopez-sepulcre2015} also contains cold sources (see brown diamond points), the authors only considered some of these points to derive their fit (see filled points in the left panel of Fig. \ref{h2co/hnco_vs_nh2cho}). The power-law fits we derive for all the observed points (red, green and brown points) are given by the equations $\rm [NH_2CHO]=32.14\times[HNCO]^{1.29}$ and $\rm [NH_2CHO]=4.65\times[H_2CO]^{1.39}$.

The left panel shows that a tight correlation between HNCO and \formamide can be drawn. However, note that hydrogenation reactions for \formamide {\it have been removed in this example}. This clearly illustrates how the correlation of HNCO and \formamide does not come from a direct chemical link between the two, but rather from the same response of the two species to environmental conditions (more precisely to temperature). The changes in gas density does not seem to drive this correlation because the L1544 core centre model has a higher density than the cold envelope of IRAS16293 but the abundance of both species is much lower in the former than in the latter. However, the temperature progressively increases from L1544 to the cold envelope of IRAS16293, to the hot corino, triggering different processes on the surface of dust grains such as radical diffusion and thermal evaporation, both important mechanisms in the formation of formamide. Note that non-thermal desorption processes are responsible for the difference between the predicted abundances of NH$_2$CHO and HNCO between the positions of the core centre and the methanol peak. 

In the right panel of Fig. \ref{h2co/hnco_vs_nh2cho}, we demonstrate that \formamide is correlated with H$_2$CO but this time the correlation is due to a chemical link between the two molecules. The latter observational trend was less clear in \citet{mendoza2014} than for HNCO because H$_2$CO is much more reactive in the gas phase than HNCO. This shows that one should be cautious over interpreting correlations between two molecules because it does not necessary imply that the two species are chemically linked (chemical correlation) but rather that they are formed through similar processes across different environments (physical correlation).

\section{Conclusions}

We have updated \uclchem with a new treatment of grain-surface reactions by adding several processes such as diffusion of molecules on the grain surface, chemical reactive desorption, and reaction-diffusion competition. With this new treatment, we implemented a new set of reactions to model the abundance of molecules with a peptide-like bonds (e.g. HNCO, NH$_2$CHO, CH$_3$NCO) in various star-forming environments ranging from cold dark cores to hot corinos found in low-mass proto-stars. We studied the impact of gas phase and grain surface chemistry in the formation of formamide and we found that hydrogenation of HNCO to form \formamide tends to overestimate its abundance. The gas-phase reaction needs to be coupled with radical-radical reactions on the grain surface to recover the observed abundance in the hot corino model. Moreover, our results show that the gas phase reaction to form formamide $\rm NH_2 + H_2CO \rightarrow NH_2CHO + H$ might have a higher activation barrier of 25 K instead of the calculated 4.88\,K. HNCO and CH$_3$NCO abundances are well reproduced and we discussed the impact of important reactions to form and destroy these two species and we give better constraints to the reaction rates of the methylation of CHNO isomers. Finally, the modelled abundance of HNCO and \formamide show a power-law correlation despite the absence of the hydrogenation of HNCO to form NH$_2$CHO, demonstrating that this correlation inherits from an environmental behaviour driven by the temperature instead of a pure chemical link.

\section*{Acknowledgements}
We acknowledge Niels Ligterink and Ian Smith for valuable suggestions and comments. D.Q. and I.J.-S acknowledge the financial support received from the STFC through an Ernest Rutherford Grant and Fellowship (proposals number ST/M004139 and ST/L004801). J.H. is funded by an STFC studentship. S.V. acknowledges a grant funded by the STFC (grant ST/M001334/1). A.C. post-doctoral grant is funded by the ERC Starting Grant 3DICE (grant agreement 336474).

\bibliographystyle{mnras}
\bibliography{All_ref} 


\appendix

\section{Chemical processes added to \uclchem}\label{chem_proc}

In this appendix we detail our updates to the treatment of the grain surface chemistry applied to the gas-grain chemical code {\sc uclchem}.

\subsection{Grain surface diffusion}

We have implemented in \uclchem the diffusion mechanism described in \citet{hasegawa1992}. This formalism is extensively used in the literature in several chemical codes \citep[e.g. \textsc{nautilus},][]{ruaud2016}. Briefly, the rate at which two species A and B can diffuse and meet on the grain surface is given by:
\begin{equation}\label{diff_eq}
	k_{\textrm{AB}} = \kappa_{\textrm{AB}}\left(k^{\textrm{A}}_{\textrm{hop}} + k^{\rm B}_{\textrm{hop}}\right)\frac{1}{N_{\textrm{site}}\,n_{\rm dust}},
\end{equation}
where $N_{\textrm{site}}$\,$\sim$\,$2\times10^6$ is the number of sites on the grain surface and $n_{\rm dust}$ is the number density of dust grains:
\begin{equation}
	n_{\rm dust} = \frac{3\,n_{\rm H}\,amu}{4\uppi\,r_{{\rm gr}}^3\,n_{{\rm gr}}\,gtd},
\end{equation}
where $r_{{\rm gr}}=0.1\,\mu$m is the grain radius, $n_{{\rm gr}}$ is the density of a dust grain assumed to be 3\,g\,cm$^{-3}$, $gtd=100$ is the gas-to-dust mass ratio, $n_{\rm H}$ is the total hydrogen number density and $amu=1.66053892\times10^{-24}$\,g is the atomic mass unit.

$k^{\textrm{X}}_{\textrm{hop}}$ is the thermal hopping rate of the species $X$ on the grain surface defined as \citep{hasegawa1992}:
\begin{equation}
	k^{\textrm{X}}_{\textrm{hop}} = \frac{1}{t_{\rm hop}} = \nu_0\,\exp\left(-\frac{E{\rm _b}}{T_{\rm gr}}\right),
\end{equation}
where $t_{\rm hop}$ is the hopping time between two grain surface sites, $T_{\rm gr}$ is the grain temperature, and $E{\rm _b}$ is the diffusion energy in K (potential energy well to be overcome between two surface sites). $E{\rm _b}$ is estimated using the binding energy of the species onto the grain surface, $E{\rm _D}$. Several values have been tested in the literature and we adopted the most commonly used of $E{\rm _b} = 0.5\,E{\rm _D}$ as suggested by recent results from \citet{minissale2016-2}. This value has also been used recently by \citet{vasyunin2017}. $\nu_0$ is the characteristic vibration frequency of species X and is calculated as \citep{hasegawa1992}:
\begin{equation}
	\nu_0 = \sqrt{\frac{2\,k_B\,n_S\,E{\rm _D}}{\uppi^2\,m}},
\end{equation}
where $k_B$ is the Boltzmann constant, $n_S$ the site density on the grain surface, and $m$ the mass of species $X$. $n_S$ is linked to $N_{\textrm{site}}$ using:
\begin{equation}
	n_S = \frac{N_{\textrm{site}}}{4\,\uppi\,r_{{\rm gr}}^2}.
\end{equation}
In {\sc uclchem}, $n_S$ is fixed to 1.5$\times$10$^{15}$\,cm$^{-2}$ and is used to calculate the value of $N_{\textrm{site}}$. The typical range value for $\nu_0$ is between 10$^{12}$ and 10$^{13}$\,s$^{-1}$.
The term $\kappa_{\textrm{AB}}$ in Eq. (\ref{diff_eq}) defines the probability for the reaction between species A and B to occur. This probability can be described using a quantum mechanical probability for tunnelling through a rectangular barrier of thickness $a$:
\begin{equation}\label{tunnelling}
	\kappa_{\textrm{AB}} = \exp{\left[-\frac{2\,a}{\hbar}\,\sqrt{\,2\,\mu\,k_B\,E{\rm _A}}\right]},
\end{equation}
where $\hbar$ is the reduced Planck constant, $\mu$ is the reduced mass, and $E{\rm _A}$ is the activation energy in K of the reaction. The  rectangular barrier thickness is typically assumed to be 1\,\AA~\citep{hasegawa1992}. However, we use a value of 1.4\,\AA~in our study as it fits better the ice composition. A similar approach has been taken by \citet{vasyunin2017}, who found that a width of 1.2\,\AA~ fitted better their results. We note that the rectangular barrier thickness value has been modified in recent studies to range between [1.0 -- 2.0]\,\AA~ \citep{garrod2011, taquet2013, vasyunin2017}.

The probability $\kappa_{\textrm{AB}}$ can also be expressed as:
\begin{equation}\label{activ_energy}
	\kappa_{\textrm{AB}} = \exp{\left[-\frac{E{\rm _A}}{T_{\rm gr}}\right]}.
\end{equation}
\uclchem automatically chooses which probability is the largest between Eqs. (\ref{tunnelling}) and (\ref{activ_energy}) and uses it in Eq. (\ref{diff_eq}).
For exothermic and barrierless reactions, $E{\rm _A}$ is null and Eqs. (\ref{tunnelling}) and (\ref{activ_energy}) give a probability $\kappa_{\textrm{AB}}$ equal to one.

\subsection{Reaction-diffusion competition}

The grain surface diffusion rate calculated above only considers that the species A and B are moving on the grain surface \textit{prior to} reacting. However, molecules and radicals may directly form in a given surface site and directly react \textit{in situ} before moving to an adjacent site. Finally, the species A and B can also evaporate from the grain surface before having the opportunity to react or diffuse.
The combination of these effects is called the reaction-diffusion competition and it has been introduced by \citet{chang2007} and \citet{garrod2011}.
To take into account this effect, we define the probabilities for the diffusion, reaction, and evaporation for species A and B \citep{chang2007}:
\begin{eqnarray}
	p_{\rm diff} &=& k^{\textrm{A}}_{\textrm{hop}} + k^{\textrm{B}}_{\textrm{hop}},\\
	p_{\rm reac} &=& max(\nu{\rm ^A_0}, \nu{\rm ^B_0})\times\kappa_{\textrm{AB}},\\
	p_{\rm evap} &=& \nu{\rm ^A_0}\,\exp\left(-\frac{E{\rm ^A_D}}{T_{\rm gr}}\right) + \nu{\rm ^B_0}\,\exp\left(-\frac{E{\rm ^B_D}}{T_{\rm gr}}\right),
\end{eqnarray}
where the result of $max(\nu^A_0, \nu^B_0)$ represents the largest value of the characteristic frequencies of species A and B \citep{garrod2011}.
The probability for the reaction to occur is then defined by \citep[e.g.][]{garrod2011, ruaud2016}:
 \begin{equation}\label{kappa_final}
	\kappa^{\rm final}_{\textrm{AB}} = \frac{p_{\rm reac}}{p_{\rm reac} + p_{\rm diff} + p_{\rm evap}}.
\end{equation}
The value of $\kappa^{\rm final}_{\textrm{AB}}$ calculated above is then used in Eq. (\ref{diff_eq}) instead of $\kappa_{\textrm{AB}}$.

\subsection{Chemical reactive desorption}

Desorption of species from the grain surface can occur thanks to various thermal and non-thermal processes such as direct and cosmic ray induced UV photons, direct cosmic rays, and H$_2$ formation, as described in \citet{holdship2017}. We have added to \uclchem the chemical reactive desorption formalism defined by \citet{minissale2016-1}. This semi-empirical formalism describes the efficiency $\eta_{CD}$ of an exothermic reaction occuring on the surface of dust grains, to release products in the gas phase. $\eta_{CD}$ depends on the binding energies of the reactants and the exothermicity of the reaction \citep{minissale2016-1, vasyunin2017}:
\begin{equation}\label{eq_CD}
	\eta_{CD} = \exp\left(-\frac{E{\rm _D}\,N_{\rm dof}}{\epsilon_{CD}\,\Delta H_R}\right),
\end{equation}
where $N_{\rm dof}$ is the degree of freedom and is defined by $N_{\rm dof} = 3\times n_{\rm atoms}$. $\Delta H_R$ is the enthalpy of the reaction and it can be derived using the following equation:
\begin{equation}
	\Delta H_R [K] = \left[\left(\sum_{\rm reac}{\Delta_{\rm f}H} - \sum_{\rm prod}{\Delta_{\rm f}H}\right) \times \frac{4184}{k_B\,N_A}\right] + E{\rm _A}.
\end{equation}
The enthalpy of formation $\Delta_{\rm f}H$ of both reactants and products is given in kcal\,mol$^{-1}$ and converted in K using the kcal to J conversion factor of 4184\,J\,kcal$^{-1}$. $N_A$ is the Avogadro number.

To be desorbed from the grain surface to the gas phase, products need to gain velocity in a direction perpendicular to the surface. Therefore, in the \citet{minissale2016-1} formalism, products need to bounce against the surface in an elastic collision. The fraction of kinetic energy $\epsilon_{CD}$ retained by the product of mass $m$ colliding with the surface with effective mass $M$ is then defined by:
\begin{equation}
	\epsilon_{CD} = \frac{(m-M)^2}{(m+M)^2}.
\end{equation}
The effective mass $M$ of the grain surface is a poorly constrained parameter. Its value has been estimated to be approximately 130\,amu (equivalent to the mass of 11 carbon atoms, \citealp{hayes2012}). This mass is larger than that of a single carbon atom because the surface is acting as a group rather than individually in the collision. \citet{minissale2016-1} suggested that a value of 120\,amu well reproduces the trend of their experimental chemical reactive desorption measurement for bare grains. \citet{vasyunin2017} varied the effective mass between 80 and 120\,amu and the resulting abundances can change by an order of magnitude, although their predicted trend does not. In order to be consistent with the studies of \citet{minissale2016-1} and \citet{vasyunin2017}, we have fixed the value of $M$ to 120\,amu.

Finally, the expression (\ref{eq_CD}) is only valid for bare grains. In the case of amorphous solid water (ASW) surfaces, \citet{minissale2016-1} derived that $\eta^{\rm ice}_{CD} = \eta^{\rm bare}_{CD}/10$. We have also used this value in {\sc uclchem}. In addition, we have adopted their constraints to the following chemical reactive desorption coefficients:  $\eta^{\rm ice}_{CD}{\rm (OH+H)} = 25\,\%$, $\eta^{\rm ice}_{CD}{\rm (O+H)} = 30\,\%$, $\eta^{\rm ice}_{CD}{\rm (N+N)} = 50\,\%$.

\subsection{Limitations of the rate equation approach}

We note that large differences can be found between the rate equation approach and stochastic methods (such as Monte Carlo models see, e.g. \citealp{cuppen2009, vasyunin2009, lamberts2014, chang2014, chang2016}). The rate equation approach considers macroscopic effects directly applied to grain surfaces which may lead to large uncertainties, especially when abundances of reactants on the grain surface are low \citep{gillespie1976, green2001, charnley2001}. However, we use the rate equation approach in {\sc uclchem} because of the convenience, stability, and the rather fast numerical performance of the code, even for reaction networks consisting of thousands of reactions involving hundreds of molecules (see review by \citealp{cuppen2017}). More complex codes (stochastic or not) can be much slower, depending on the complexity of the chemistry (bulk chemistry, 2D, 3D...) taking typically days or weeks to run compared to the minutes required for the rate equation approach.


\bsp	
\label{lastpage}
\end{document}